\documentclass[aps, floatfix]{revtex4}

\usepackage{graphicx}
\usepackage{epsfig}
\usepackage{amssymb}

\begin{document}
\newcommand{\be}{\begin{eqnarray}}
\newcommand{\ee}{\end{eqnarray}}
\newcommand\del{\partial} 
\newcommand\nn{\nonumber}
\newcommand{\Tr}{{\rm Tr}}
\newcommand{\mat}{\left ( \begin{array}{cc}}
\newcommand{\emat}{\end{array} \right )}
\newcommand{\vect}{\left ( \begin{array}{c}}
\newcommand{\evect}{\end{array} \right )}
\newcommand{\tr}{\rm Tr}


\title{QCD in One Dimension  at Nonzero Chemical Potential}

\author{L. Ravagli}
\affiliation{Cyclotron Institute and Physics Department TEXAS A\&M University, College Station, 
Texas 77843-3366, USA}
\author{J.J.M. Verbaarschot}
\affiliation{Niels Bohr International Academy, Blegdamsvej 17, DK-2100, Copenhagen {\O}, Denmark,}
\affiliation{Niels Bohr Institute, Blegdamsvej 17, DK-2100, Copenhagen {\O}, Denmark}
\affiliation{Department of Physics and Astronomy, SUNY, Stony Brook,
 New York 11794, USA}


\begin{abstract}
Using an integration formula recently derived by Conrey, Farmer
and Zirnbauer, we calculate the expectation value of the phase factor of the
fermion determinant for the staggered lattice QCD action in 
one dimension. We show that the chemical potential
can be absorbed into the quark masses; the theory is in the same chiral symmetry
class as QCD in three dimensions at zero chemical potential. 
In the limit of a large number of colors and
 fixed number of lattice points, chiral symmetry is broken
spontaneously, and our results are in agreement
with expressions based on a chiral Lagrangian. In this limit,
the eigenvalues of the Dirac operator 
are correlated according to random matrix theory for QCD in three dimensions.
The discontinuity
of the chiral condensate is due to an alternative to the Banks-Casher formula
recently discovered for QCD in four dimensions at nonzero chemical potential.
The effect of temperature on the average phase factor is discussed
in a schematic random matrix model.

\end{abstract}

\maketitle
\setlength{\textwidth}{15cm}

\section{Introduction}

In spite of significant recent progress, 
QCD at nonzero chemical potential remains  a notoriously 
hard problem \cite{kimrev,schmidtrev,mariarev}. In
particular, first principle nonperturbative results at low temperature 
are absent because
the phase of the fermion determinant invalidates  probabilistic methods to
evaluate the partition function. This problem is known as the sign problem.
The sign problem is particularly severe 
if the phase of the fermion determinant
results in a different free energy, i.e. 
\be
\frac{  \langle {\det}^{N_f} D \rangle} 
{\langle |{\det}^{N_f} D| \rangle} = e^{V(F_{\rm pq} -F_{N_f})}\qquad {\rm with} \qquad
F_{N_f} > F_{\rm pq}.
\label{eq1}
\ee 
If the phase quenched free energy, $F_{\rm pq}$ differs from the free
energy of QCD with $N_f$ flavors, $F_{N_f}$, the number of required gauge field 
configurations grows exponentially with the volume. The ratio defined
in equation (\ref{eq1}) can also be interpreted as the average phase
factor of the fermion determinant \cite{SVphase1,SVphase2}
\be
\langle e^{iN_f \theta}\rangle_{\rm pq} =
\frac{  \langle  e^{iN_f\theta} | {\det}^{N_f} D| \rangle} 
{\langle |{\det}^{N_f} D| \rangle }.
\label{eq2}
\ee 
Alternatively, one can define the average phase factor with respect to the 
full QCD partition function \cite{SVphase1,SVphase2}
\be
\langle e^{2i \theta}\rangle_{ N_f} =
\frac{  \langle  e^{2i\theta}  {\det}^{N_f} D\rangle} 
{\langle {\det}^{N_f} D \rangle },
\label{eq3}
\ee 
which has  also been used in the literature
\cite{gibbs1,Toussaint,deFL,NakamuraPhase,Ejiri,schmidt,Allton,SVphase1,SVphase2}
as a measure for the severity of
the sign problem.

When the chemical potential is sufficiently small, the average phase
factor can be studied by means
of chiral perturbation theory \cite{SVphase1,SVphase2}. 
Recently, the average phase factor was analyzed
in the microscopic domain of QCD \cite{SV,V}, 
where only the constant fields in the chiral
Lagrangian contribute to the mass and chemical potential dependence of the
partition function. It was found that the sign problem is not serious for
$\mu< m_\pi/2$. For $\mu >m_\pi/2$ the chiral condensate of the
phase quenched theory rotates into a pion condensate resulting 
in  a free energy
that is different from the full theory  and a severe sign problem.

 In this paper we study the sign problem for Euclidean QCD in one dimension.
In one dimension, the only effect of the gauge field is in the boundary
conditions, and lattice QCD simplifies to a matrix integral. A general formula
for exactly this type of integrals was recently derived by Conrey,
Farmer and Zirnbauer \cite{CFZ}. Using what we will call the 
CFZ formula, exact analytical
results for the  one dimensional QCD partition function and
the average phase factor will be obtained.

The study of QCD in one dimension at finite chemical potential has 
had a long history. One reason to study this model is that it is an effective
model for the large $N_c$ limit of strong 
coupling QCD 
\cite{ilgenfritz,damgaard,gocksch-ogilvie,hochberg,bilic-new,kim}. Among others,
it has been successfully  used to explain \cite{gibbs-plb,gocksch,bilic}  
puzzling lattice results for quenched QCD at nonzero chemical potential.
More recently, one dimensional  QCD was used to study the complex zeros 
of the partition function \cite{gupta1d,maria1d,mariarev}.

The sign problem of QCD at nonzero chemical potential arises because 
of the nonhermiticity of the Dirac operator. 
QCD in one dimension is {\it not} truly nonhermitean:
Instead of being scattered in a two-dimensional domain of the 
complex plane \cite{all},
the eigenvalues of the Dirac operator are localized on an ellipse in the complex
plane \cite{gibbs-plb,gibbs1}. Another manifestation of the mild nonhermiticity
is that the chemical potential can be absorbed into the quark masses.
For each
flavor in one dimension with quark mass $m$, we can associate flavors with mass
$m+\mu$ and $-m +\mu$. In the
large $N_c$ limit, this results in spontaneous chiral symmetry breaking
according to $U(2N_f) \to U(N_f)\times U(N_f)$. Therefore,
QCD in one dimension  is in the same chiral symmetry class as QCD in three dimensions. 
The microscopic limit of QCD in one dimension is
equivalent to the microscopic
limit of QCD in three dimensions \cite{jvzahed3}. 
Another peculiarity of QCD in one dimension is that, already in
the free theory, the eigenvalues near zero are spaced  inversely proportional to the volume. 
Therefore, in the thermodynamic limit (i.e. at zero temperature) the chiral condensate is discontinuous 
across the imaginary axis. However, this type of symmetry breaking is not a collective phenomenon,
and there are no associated Goldstone bosons. It is reminiscent to the alternative to 
Goldstone's theorem proposed by McKane and Stone \cite{stone}.
Spontaneous symmetry breaking with Goldstone bosons
takes place for a fixed number of lattice points in the limit $N_c \to \infty$.

A chemical potential excites color singlet excitations 
with baryon charge $q_k$ and mass $M_k$
when $q_k \mu > M_k$. This is also the case for one-dimensional QCD. For 
gauge group $U(N_c)$, the only color singlet excitations are mesons that are
uncharged with respect to the chemical potential, whereas
for $SU(N_c)$ gauge group we have both neutral 
mesonic and charged baryonic color singlet
excitations. The complex conjugate of the fermion determinant can be interpreted
in terms of conjugate quarks \cite{gocksch,misha} which have a baryon charge that
is opposite to that of regular quarks. Therefore, both for $U(N_c)$ and $SU(N_c)$,
 the phase quenched partition function, where the fermion determinant has
been replaced by its absolute value,  
has charged mesonic excitations 
made out of quarks and conjugate anti-quarks. This results in a phase
transition at $\mu = \mu_c=m_\pi/2$. The $SU(N_c)$ theory with $N_f$ flavors
has only charged baryonic excitations and will have a phase transition at 
$\mu = \mu_c = m_B/N_c$. In one dimension it turns out that 
the pion and the baryon have the same mass per quark number so that 
the critical chemical potential of the phase quenched $U(N_c)$ theory
and the $SU(N_c)$ theory is the same. 

For the $U(N_c)$ partition function we expect the sign problem to be severe
when $\mu > \mu_c$ because the phase quenched partition function has
a phase transition at $\mu=\mu_c$  
whereas the normal theory remains in the same phase.
For $SU(N_c)$, both the normal partition function
and the phase quenched partition function do have a phase transition  
to a phase of free quarks at
$\mu=\mu_c$ and have the same free energy  not only for
 $\mu<\mu_c$ but also for $\mu > \mu_c$.
Therefore, in one dimension there is no severe sign problem
for  $SU(N_c)$ QCD.

Because one dimensional QCD with gauge group $U(N_c)$ does not have a phase transition, the
chiral condensate in the thermodynamic limit or large $N_c$ limit
is discontinuous at $m=0$, independent of the value of the chemical potential.
Because the Dirac eigenvalues are located on an ellipse for $ \mu \ne 0$, 
this discontinuity cannot
be related to the Dirac spectrum by means of the Banks-Casher formula. In \cite{OSV} a
different mechanism to explain the discontinuity was discovered . It was found that 
an oscillating contribution to  
the spectral density with an amplitude that diverges exponentially
with the volume, is responsible for the discontinuity of the chiral condensate. We will
show that a similar mechanism is at work for $U(N_c)$ gauge theory in one dimension.

Lattice QCD in one dimension will be introduced in section II,  where we also discuss its
continuum limit, mean field results and the Conrey-Farmer-Zirnbauer formula. In section
III we will evaluate the $U(N_c)$ partition function and its phase quenched version. The
average phase factor for $U(N_c)$ is calculated in section IV. In section V we study one
dimensional QCD with $SU(N_c)$ as gauge group and evaluate the regular and phase quenched
partition functions and the average phase factor. The effect of temperature will
be illustrated with results from a schematic random matrix model in section VI.
The connection with the microscopic domain
of QCD in three dimensions 
is discussed in section VII, and the Dirac spectrum and its relation
with chiral symmetry breaking is analyzed in section VIII. Concluding remarks are made in
section IX.

\section{QCD in One Dimension}

\subsection{Lattice QCD}
The staggered lattice QCD Dirac operator in one dimension is given by
\be
D=\left( \begin{array}{cccc} 
mI & e^{\mu}U_{12}/2 & {\ldots} &  e^{-\mu}U_{n1}^{\dagger}/2\\
-e^{-\mu}\,U_{12}^{\dagger}/2 & mI & \cdots  &0\\ 
\vdots& &&\vdots\\
 0& \cdots&mI&e^{\mu}U_{n-1\,n}/2\\
-e^{\mu}U_{n1}/2& \cdots &-e^{-\mu}U_{n-1\,n}^{\dagger}/2&mI\\
\end{array} \right).
\label{dirac}
\ee
We have used anti-periodic boundary conditions and the gauge
fields on the links are taken to be in $U(N_c)$ or $SU(N_c)$.
The chemical potential $\mu$ is an imaginary vector field introduced according to the 
Hasenfratz-Karsch prescription \cite{HK}.
In general, the lattice QCD partition function for $N_f$ flavors with mass $m$
is given by
\be
Z_{N_f}(\mu_c,\mu)= \int \prod_l dU_l {\det}^{N_f} D\, e^{-S_{\rm YM}/g^2},
\ee
where the product is over the links and $S_{\rm YM}$ is the
Yang-Mills action. In one dimension, the partition function simplifies
substantially. First, there is no Yang-Mills action, and second,
by unitary transformations the integral over the links can be
reduced to a single integral. In the gauge where all gauge 
fields except  $U_{n\,1}\equiv U$ are equal the unity, 
the fermion determinant reduces to \cite{bilic}
\be
\det D =  2^{-nN_c}\det[e^{n\mu_c} + e^{-n\mu_c}+ e^{n \mu} U +e^{-n \mu} U^\dagger],
\label{det1}
\ee
with $\mu_c$  given by
\be
\mu_c = \sinh^{-1} m. 
\ee
This value will turn out to be the critical value of the chemical potential. From now on, the overall
factor $2^{-nN_c}$ will be absorbed into the normalization of the partition function. To avoid
sign factors, $n$ is taken to be even throughout this paper.

The standard method to evaluate the QCD partition function in one dimension is 
to use  the eigenvalues
of $U$ as integration variables \cite{damgaard,gocksch-ogilvie,ilgenfritz,gibbs1,bilic}.
For example, one finds this way that the  result for $N_f = 1$ with 
$U(N_c)$ as gauge group is  
given by \cite{bilic}
\be
Z_{N_f=1}(\mu_c,\mu) = \int_{U(N_c)} dU \det D = \frac{\sinh((N_c+1)n\mu_c)}{\sinh(n\mu_c)}.
\label{eq8}
\ee
For most partition functions that are considered in this paper it is not possible to obtain analytical
results by means of this method. Instead we will use powerful 
integration formulas for unitary integrals that were recently derived in 
\cite{CFZ}. These integrals are based on the 
color-flavor transformation \cite{color-flavor} which
has also been applied to lattice QCD with baryons 
in the canonical ensemble \cite{slit,zsu}.

\subsection{Continuum Theory}

\label{sec:continuum}
The continuum limit of the staggered lattice action is given by
\be
D_{\rm cont} = \mat m & \del_0 + iA_0 + \mu \\ 
\del_0 + iA_0 + \mu  & m \emat,
\label{dirac1d}
\ee
where $A_0$ is a Hermitean $N_c\times N_c$ matrix and the off-diagonal blocks
connect even and odd lattice sites. What is special in one
dimension 
is that the off-diagonal blocks are identical. 
The eigenvalue equation $(\del_0 + iA_0)\psi_k = iu_k \psi_k $ is solved by
\be
\psi = P e^{-i\int_0^t A_0 dt}  \chi_0 \qquad
{\rm with} \qquad  \del_0 \chi_0 = i E \chi_0 ,
\ee
with $P e$ the path ordered exponent. 
Nontrivial eigenvalues are obtained by imposing boundary
conditions on $\psi$. At nonzero chemical potential
the eigenvalues of the Dirac operator are given by
\be
\lambda_k = m \pm (iu_k+ \mu) \qquad {\rm with} \qquad u_k \in \mathbb{R}.
\ee
This is very different from QCD with $d\ge 2$ where the eigenvalues
of the QCD Dirac operator at $\mu \ne 0$ are scattered in the complex plane.
Also the eigenvalues of the one dimensional staggered lattice Dirac 
at nonzero chemical potential
are localized on a curve in the complex plane.

Another consequence of  the structure of the Dirac operator
(\ref{dirac1d}) is that
the fermion determinant can be rewritten as
\be
\det D_{\rm cont} = \det (\del_0 +iA_0 +\mu +m) \det(\del_0 + iA_0  +\mu -m),
\label{dirac3}
\ee
which can be interpreted as  
a two-flavor partition function with masses $m +\mu$ and $-m+\mu$.
This Dirac operator has the same structure as QCD in three dimensions:
In the large $N_c$ limit,  chiral symmetry is broken
spontaneously according to 
$U(2) \to U(1) \times U(1)$ with the squared Goldstone masses 
proportional to the difference of the positive and negative 
quark masses, i.e. $m_\pi^2 \sim m -(-m)=2m$.
At low energy, these Goldstone modes interact according to a chiral Lagrangian
determined by the pattern of chiral symmetry breaking. In the large $N_c$ 
limit and $m_\pi\beta \ll 1$ (with $\beta$ the length of the one-dimensional box)
the QCD partition function in one dimension is therefore equivalent to the low-energy
limit of QCD in three dimensions.
In section  \ref{qcd3} this will be worked out explicitly  for the microscopic
limit of the partition function. This is the limit 
\be
\rho(0) \to \infty \quad {\rm with} \quad
m \pi\rho(0)= {\rm fixed} \quad {\rm and} \quad \mu\pi \rho(0) ={\rm fixed},
\label{micro}
\ee
where $\rho(0)$ is density  of the eigenvalues, or their projections onto
the imaginary axis, close to zero. 
Since $\rho(0)= nN_c/2\pi$  (see section VIII),
this is the limit $n N_c\to \infty$ with $n N_c m$
and $n N_c \mu$ fixed. When we use the term microscopic limit, we always mean the
universal microscopic limit. This is the limit (\ref{micro}) associated with the formation
of Goldstone bosons, i.e. the limit $N_c\to \infty$ at fixed $n$.

\subsection{Mean Field Limit for Large $N_c$}

\label{largen}
In the large $N_c$ limit, chiral symmetry is broken spontaneously even in
one dimension, so that its low-energy limit is a theory of Goldstone bosons.
In this section we give general arguments that determine
the chemical potential dependence of the partition function 
in the microscopic domain where the Compton wavelength of the 
Goldstone bosons is much larger than $\beta$.

As was argued in \cite{SVphase1, SVphase2}, in the microscopic
domain, the mean field limit of the partition function is given by
\be
Z = J \left ( \prod_k \frac 1{m_\pi(\mu)} \right ) e^{-V F},
\label{mftform}
\ee
where $J$ is the value of the integration measure at the saddle point, $F$
is the free energy, $V$ is the space time volume, and $m_\pi$ are the masses of
the Goldstone bosons. Let us apply this result to the average phase factor in
one dimension.
For $\mu < \mu_c $, in the limit $T\to 0$, only the vacuum state contributes to
the partition function, so that the free energy is independent
of $\mu$.  For quark mass $m$, the equivalent 
QCD$_3$ mass matrix of the Dirac operator $D$ in (\ref{dirac3}) 
is given by ${\rm diag}(-m+\mu, m+\mu)$, 
whereas the hermitean
conjugate Dirac operator $D^\dagger$, has ${\rm diag}(-m-\mu, m -\mu)$ as  
equivalent QCD$_3$ mass matrix.
For $N_f$ flavors, the average phase factor 
is the partition function with $N_f+1$ fermionic quarks and one conjugate bosonic quark.
We thus have $2(N_f+1)^2$ Goldstone bosons made out of two fermionic quarks
with squared mass $m_\pi^2 = 2 m G$ (with $G$ a constant),
$4(N_f+ 1)$ fermionic Goldstone modes, half of them with squared mass
$m_\pi^2 = 2(m-\mu)G$, and the other half with squared mass $m_\pi^2=2(m+\mu)G$.
Finally, we have 2 Goldstone bosons composed out of two bosonic quarks
with squared mass equal to $ m_\pi^2 = 2m G$.  The fermionic partition function $Z_{N_f}$ has
$2N_f^2$ Goldstone bosons all with squared mass $m_\pi^2= 2mG$. Using (\ref{mftform})
we thus find
\be
\langle e^{2i\theta} \rangle_{N_f}  = \frac{Z_{N_f+1|1^*}(\mu_c,\mu)}{Z_{N_f}(\mu_c,\mu)}=
\left ( 1 - \frac {\mu^2}{m^2} \right )^{N_f+1} \quad {\rm for}
\quad \mu < \mu_c.
\ee

For phase quenched partition functions with $N_f$ flavors and $N_f$
conjugate flavors, the equivalent QCD$_3$ symmetry breaking pattern is
$U(4N_f) /(U(2N_f)\times U(2N_f))$, so that the total number of Goldstone
bosons is equal to $8N_f^2$. Of these,
$4N_f^2$ Goldstone bosons have squared mass $m_\pi^2=2mG$, half made out of two quarks and
the other half out of two conjugate quarks. The other $4N_f^2$ Goldstone
bosons are composed out of a quark and a conjugate anti-quark quark, half with squared mass
$m_\pi^2= 2(m-\mu)G$ and the other half with squared mass $m_\pi^2=2(m+\mu)G$. 
All equivalent QCD$_3$ Goldstone 
bosons of  the full QCD
partition function with $2N_f$ flavors  have a squared mass  equal to $m_\pi^2=2mG$. 
For $\mu<\mu_c$ the free energy of the 
phase quenched partition function and the full QCD partition function is
the same so that, using (\ref{mftform}), the 
phase quenched average phase factor is given by
\be
\langle e^{2i\theta} \rangle_{N_f+N_f^*} = \frac{Z_{N_f+1+(N_f-1)^*}(\mu_c,\mu)}{Z_{2N_f}(\mu_c,\mu)} =
\left(1-\frac{\mu^2}{m^2}\right)^{N_f^2} \quad {\rm for} \quad \mu<\mu_c.
\label{phmftpq}
\ee
 
Below we will show that the results derived in this section also follow
from the
zero temperature microscopic limit of the exact evaluation of the partition function.

\subsection{The Conrey-Farmer-Zirnbauer  formula}

Exactly the integrals that are required for the evaluation of the average
phase factor of QCD at nonzero chemical potential in one dimension were
studied in a recent paper by Conrey, Farmer and Zirnbauer \cite{CFZ}. 
They considered the partition function
\be
Z(\{\psi_k,\phi_k\})=\int_{U(N_c)}dU
\prod_{j=1}^p \frac {\det(1-e^{\psi_j}U)}{\det(1-e^{\phi_j}U)}
\prod_{l=p+1}^{p+q} \frac {\det(1-e^{-\psi_l}U^\dagger)}{\det(1-e^{-\phi_l}U^\dagger)},
\ee
with $dU$ the Haar measure of $U(N_c)$ and $\psi_k$, $\phi_k$ complex parameters with ${\rm Re}(\phi_j) < 0 < {\rm Re}(\phi_l)$.
Using the color flavor transformation \cite{color-flavor} and Howe's
theory of supersymmetric dual pairs, they derived the following formula
\be
Z(\{\psi_k,\phi_k\})=
\sum_{\pi\in S_{p+q}/(S_p\times S_q)}
\prod_{l+1}^{p+q} e^{N_c(\pi(\psi_l)-\psi_l)}
\prod_{j=1}^p\frac{ (1-e^{\phi_j-\pi(\psi_l)})(1-e^{\pi(\psi_j) -\phi_l})} 
{ (1-e^{\pi(\psi_j)-\pi(\psi_l)})(1-e^{\phi_j -\phi_l})} .
\label{cfz}
\ee
The sum is over permutations in $S_{p+q}/(S_p \times S_q)$ that 
interchange any of the $\psi_1, \cdots, \psi_p$ with any
of the $\psi_{p+1},\cdots, \psi_{p+q}$.

Partition functions with an unequal number of bosonic and fermionic determinants
can be obtained from special limits of (\ref{cfz}). In the case of only
fermionic determinants, an equivalent expression was first obtained
in \cite{snaith,forrester}. Expressions for degenerate parameters can be
derived by carefully taking limits of the above formula.

Unitary matrix integrals can also be 
calculated by using an eigenvalue
representation of the unitary matrices. The integrals we are interested
in are of the form
\be
Z= \int_{U(N_c)} dU \prod_k F(e^{i\theta_k}),
\ee
where $\exp(i\theta_k)$ are the eigenvalues of $U$. 
Using an eigenvalue representation
of the unitary matrices, the orthogonal polynomial method can be used
to express $Z$ as \cite{damgaard}
\be
Z  = \det(B_{k-l})_{k,l = 0,\cdots, N_c-1}, \label{op-form}
\ee
where
\be
B_k = \frac 1{2\pi} \int_{-\pi}^\pi d\theta e^{ik\theta} F(\{e^{i\theta}\}). 
\ee
This is the method that was used in the literature on 
one-dimensional QCD prior to this
paper \cite{damgaard,gocksch-ogilvie,ilgenfritz,gibbs1,bilic}.  
In a few cases the determinant in (\ref{op-form}) could be evaluated explicitly 
resulting in expressions that are similar to those derived directly from the CFZ-formula.
We have used (\ref{op-form}) to numerically check the results obtained
by means of the CFZ-formula.

\section{Exact Evaluation of the One-Dimensional  $U(N_c)$ Partition Function}

\subsection{Partition Function for arbitrary $N_f$}

In this section we evaluate the one-dimensional $U(N_c)$ QCD partition function for 
$N_f $ flavors.

To apply the CFZ formula (\ref{cfz})    we
rewrite the determinant (\ref{det1}) as
\be
\det D= e^{\mu_cnN_c} \det(1+e^{n(\mu-\mu_c)} U)\det(1+e^{n(-\mu-\mu_c)} U^\dagger).
\ee

For arbitrary $N_f$ we then find the remarkably simple
answer
\be
Z_{N_f}(\mu_c,\mu)\equiv \int_{U(N_c)}dU {\det}^{N_f} D = \sum_{\sigma \in S_{2N_f}/S_{N_f}\times S_{N_f}}
\prod_{k=1}^{N_f} \prod_{l=1}^{N_f} \frac{e^{N_c  m_{\sigma(+\, k)}}}
{1-\exp(m_{\sigma(-\,l)} - m_{\sigma(+\,k)})}, 
\label{zirn}
\ee
with
\be
m_{-\, k} = - n\mu_{c\, k} \qquad {\rm and} \qquad 
m_{+\, k} =  n\mu_{c\, k}.
\ee
The sum is over all permutations that interchange positive and negative
masses. Notice that the $\mu$-dependence has canceled from this expression.
This also follows from an expansion of the determinant in powers of $U$ and
$U^\dagger$. Only 
terms with an equal number of factors $U$ and factors $U^\dagger$ are
non-vanishing.

Although the result for degenerate positive and negative masses can be obtained by carefully
taking limits of (\ref{zirn}),
it is simpler to start
from a different representation 
\cite{snaith,forrester} of partition function (\ref{zirn}) given by
\be
  Z_{N_f}(\mu_c,\mu)&=& \frac {1}{{\prod}_{1\le k<l\le 2N_f}(e^{M_l}-e^{M_k})}
\left | \begin{array}{ccccccc}
 1 & e^{m_{-\,1}}& \cdots & e^{(N_f-1)m_{-\,1}} &e^{(N_c +N_f)m_{-\,1}} &\cdots &
e^{(N_c +2N_f-1)m_{-\,1}}\\
\vdots & \vdots & &\vdots & \vdots & & \vdots \\
 1 & e^{m_{-\,N_f}}& \cdots & e^{(N_f-1)m_{-\,N_f}} &e^{(N_c +N_f)m_{-\,N_f}} &\cdots &
e^{(N_c +2N_f-1)m_{-\,N_f}}\\
 1 & e^{m_{+\,1}}& \cdots & e^{(N_f-1)m_{+\,1}} &e^{(N_c +N_f)m_{+\,1}} &\cdots &
e^{(N_c +2N_f-1)m_{+\,1}}\\ 
\vdots & \vdots & &\vdots & \vdots & & \vdots \\
 1 & e^{m_{+\,N_f}}& \cdots & e^{(N_f-1)m_{+\,N_f}} &e^{(N_c +N_f)m_{+\,N_f}} &\cdots &
 e^{(N_c +2N_f-1)m_{+\,N_f}}
\end{array} \right |.
\;\hspace*{0.5cm}
\label{conreyform}
\ee
For convenience we have introduced the mass matrix
$M_k = (m_{1\,1},\cdots, m_{-\,N_f},m_{+\,1},\cdots,m_{+\,N_f})$.
To obtain an expression for degenerate masses, we Taylor expand the 
exponential functions $\exp(m_{+\, k})$ and $\exp(m_{-\,k})$ 
to order $N_f-1$ about 
$\exp(-n\mu_c)$ and $\exp(n\mu_c)$, respectively, and write the resulting matrix
as the product of two matrices, one containing the Taylor coefficients, and
the other containing powers of the expansion parameters. The determinant
of the second matrix can be written as a Vandermonde determinant which cancels
against part of the prefactor in (\ref{conreyform}). Our final expression
for the partition function with degenerate masses is given by
\be
&&Z_{N_f}(\mu_c,\mu) = \frac 1{(\prod_{k=0}^{N_f-1} k!)^2} 
\frac 1{(e^{n\mu_c} - e^{-n\mu_c})^{N_f^2}}\nn
\\
&&\times \left| \begin{array}{cccccc}
1  &  \cdots & 0 & 1 & \cdots & 0 
\\
e^{-n\mu_c} & \cdots &e^{-n\mu_c}
& e^{n\mu_c} &\cdots & e^{n\mu_c} 
\\ 
e^{-2n\mu_c}  & \cdots &\delta_-^{N_f-1}e^{-2n\mu_c}
&e^{2n\mu_c} &\cdots & \delta_+^{N_f-1} e^{2n\mu_c} 
\\ 
\vdots & & \vdots \vdots &  & \vdots 
\\
e^{-n(N_f-1)\mu_c} 
& \cdots &\delta_-^{N_f-1} e^{-n(N_f-1)\mu_c}
& e^{n(N_f-1)\mu_c}  
&\cdots & \delta_+^{N_f-1} e^{n(N_f-1)\mu_c} 
\\ 
e^{-n(N_c+N_f)\mu_c} 
& \cdots &\delta_-^{N_f-1} e^{-n(N_c+N_f)\mu_c}
&  e^{n(N_c+N_f)\mu_c}  
&\cdots & \delta_+^{N_f-1} e^{n(N_c+N_f)\mu_c} 
\\ 
\vdots & & \vdots \vdots  && \vdots 
\\
e^{-n(N_c+2N_f-1)\mu_c} 
& \cdots &\delta_-^{N_f-1} e^{-n(N_c+2N_f-1)\mu_c}
& e^{n(N_c+2N_f-1)\mu_c}  
&\cdots & \delta_+^{N_f-1} e^{n(N_c+2N_f-1)\mu_c} 
\\ 
\end{array}\right |,
\ee
with
\be
\delta_- = \frac d{d(-n \mu_c)},\qquad
\delta_+ = \frac d{d(n \mu_c)}.
\ee
The $k$'th column (with $k \le N_f$) 
is given by the $\delta_-^{(k-1)}$ derivative of the first
column, and the $N_f+k$'th column is given by the $\delta_+^{(k-1)}$ 
derivative of
the $N_f$'th column.

From this result one can easily derive explicit expressions for small values of $N_f$.
 The partition function for $N_f =2$ reads
\be
Z_{N_f = 2}(\mu_c,\mu)=\frac{(e^{n(N_c+2)\mu_c} - e^{-n(N_c+2)\mu_c})^2}
{(e^{n\mu_c} - e^{-n \mu_c})^4}
- \frac{(N_c+2)^2}{(e^{n\mu_c} - e^{-n \mu_c})^2}.\hspace*{0.5cm}
\label{znf2}
\ee 
The microscopic limit of this partition function is given by
\be
Z_{N_f=2}^{\rm micro}(\mu_c,\mu) = \frac{(e^{N_c \mu_c}-e^{-N_c \mu_c})^2}{16 \mu_c^4}. 
\label{micro2}
\ee

For $N_f = 3$ the partition function is given by: 
\be
Z_{N_f = 3}(\mu_c,\mu)&=&
\frac{(e^{n(N_c+3)\mu_c} - e^{-n(N_c+3)\mu_c})^3}{(e^{n\mu_c} - e^{-n \mu_c})^9}
-(N_c+3)^3
\frac{(e^{n(N_c+1)\mu_c} - e^{-n(N_c+1)\mu_c})}{(e^{n\mu_c} - e^{-n \mu_c})^7}
\\&&+N_c(N_c+2)^2
\frac{(e^{n(N_c+3)\mu_c} - e^{-n(N_c+3)\mu_c})}{(e^{n\mu_c} - e^{-n \mu_c})^7}
-\frac 14(N_c+2)^2(N_c+3)^2
\frac{(e^{n(N_c+3)\mu_c} - e^{-n(N_c+3)\mu_c})}{(e^{n\mu_c} - e^{-n \mu_c})^5}.
\nn \\
\ee

\subsection{The Phase Quenched Partition Function}

To calculate the average phase factor according to the definition (\ref{eq2}) for $N_f=2$,
we also need the one-dimensional    phase quenched QCD partition function for
two flavors which will be evaluated in this subsection.
It is defined by
\be
Z_{\rm 1+1^*}(\mu_c,\mu) &=& \int_{U\in U(N_c)} dU  \det D \det D^\dagger\\
&=&e^{2\mu n N_c} \int_{U\in U(N_c)} dU 
\det(1-e^{n(\mu-\mu_c)} U)\det(1-e^{n(-\mu-\mu_c)} U^\dagger)
\det(1-e^{n(\mu-\mu_c)} U^\dagger )\det(1-e^{n(-\mu-\mu_c)} U).\nn
\label{zpq}
\ee
This partition function is easily evaluated using the integration formulae
of \cite{CFZ}. We find
\be
Z_{1+1^*}(\mu_c,\mu) &=&
\frac{ \cosh(2 n(N_c+2)\mu_c)}{8\sinh(n(\mu_c-\mu)) \sinh(n(\mu+\mu_c)) 
\sinh^2(n\mu_c)}
+\frac{ \cosh(2 n(N_c+2) \mu))}{8\sinh(n(\mu-\mu_c))\sinh(n(\mu+\mu_c))
\sinh^2(n\mu)} \nn \\&&
+\frac 1{8  \sinh^2(n\mu) \sinh^2(n\mu_c)}.
\ee
For $\mu=0$ this result agrees with the $N_f =2 $ partition 
function given in (\ref{znf2}).
Its microscopic limit simplifies to
\be
Z_{\rm 1+1^*}^{\rm micro}(\mu_c,\mu) = \frac{e^{2nN_c \mu_c}+e^{-2nN_c \mu_c}}
{16n^4(\mu_c^2-\mu^2)\mu_c^2}
+\frac 2{16n^4\mu^2\mu_c^2} 
-\frac{e^{2nN_c \mu}+e^{-2nN_c \mu}}
{16n^4(\mu_c^2-\mu^2)\mu^2}.
\label{zpqmicro}
\ee
In section \ref{qcd3} we will show that this result is equal to
the microscopic limit of the QCD$_3$ partition function with masses
$-(\mu_c +\mu), \,-(\mu_c-\mu), \, \mu_c-\mu, \,  
\mu_c +\mu$. 

 The large $N_c$ limit of the phase quenched partition function is 
 given by 
\be
Z_{1+1^*}(\mu_c,\mu ) &=& \frac {e^{2n(N_c+2) \mu_c}}{16\sinh(n(\mu_c-\mu))
\sinh(n(\mu+\mu_c))(\sinh(n\mu_c))^2} \quad{\rm for} \quad \mu <\mu_c,\nn \\
Z_{1+1^*}(\mu_c,\mu ) &=& \frac {e^{2n(N_c+2) \mu}}{16\sinh(n(\mu-\mu_c))
\sinh(n(\mu+\mu_c))(\sinh(n\mu))^2} \quad{\rm for} \quad \mu >\mu_c,\nn \\
Z_{1+1^*}(\mu_c,\mu ) &=& \frac{N_c e^{2n(N_c+2) \mu_c}}
{8\sinh^2 (n\mu_c) \sinh(2n\mu_c)}\left[1+O(N_c^{-1}) \right ]
\quad {\rm for} \quad \mu =\mu_c,
\label{zpqlnc}
\ee
which will be used to calculate the phase quenched average phase
factor in this limit.

\section{Average Phase Factor for $U(N_c)$}

In this section we will evaluate the average phase factor, first from the
ratio of the full QCD partition function and the phase quenched partition
function  and in the next subsection starting from the definition in Eq. (\ref{eq3}).

\subsection{Phase Quenched Average Phase Factor}

For $N_f =2$, the average phase factor with the absolute value of the 
fermion determinant as weight is given by
\be
\langle e^{2i \theta} \rangle_{\rm pq} = 
\frac{ \langle e^{2i\theta} |\det (D)|^2  \rangle} {\langle |\det(D)|^2 \rangle} 
=\frac {Z_{N_f=2}(\mu_c,\mu)}{Z_{1+1^*}(\mu_c,\mu)}.
\ee 
The two-flavor partition function was evaluated in subsection III A, and 
the phase quenched partition function was calculated in subsection  III B.
The large $N_c$ limit of the phase quenched average phase factor is given by
\be
 \langle e^{2i\theta} \rangle_{\rm pq}
=\left \{ \begin{array}{ccc}
\frac{\sinh(n(\mu_c-\mu)) \sinh(n(\mu_c+\mu))}{\sinh^2(n\mu_c)}
 &\qquad {\rm for} \qquad & \mu <\mu_c, \\ && \\

\frac{\sinh^2(n\mu)\sinh(n(\mu-\mu_c))\sinh(n(\mu+\mu_c))}
{\sinh^4(n \mu_c)} e^{-2n(N_c+2)(\mu-\mu_c)}
 &\qquad {\rm for} & \mu> \mu_c,\\ && \\ 
\frac {2\cosh(n\mu_c)}{N_c \sinh(n\mu_c)}
 &\qquad {\rm for} & \mu= \mu_c  .
\end{array} \right . 
\label{phlgnu}
\ee

The large $\mu_c N_c,\, \mu N_c$ limit of the microscopic 
phase quenched partition function (\ref{zpqmicro})
for $\mu < \mu_c$ is given by
\be
Z_{\rm pq}(\mu_c,\mu) = \frac{e^{2nN_c \mu_c}}
{16n^4(\mu_c^2-\mu^2)\mu_c^2},
\label{zthpq}
\ee
whereas the two-flavor partition function in this limit reads
\be
Z_{N_f=2}(\mu_c,\mu) = \frac{e^{2nN_c\mu_c}}{16 n^4\mu_c^4},
\label{znftlg}
\ee
so that the average phase factor simplifies to
\be
\langle e^{2i\theta} \rangle_{\rm pq} = 
1- \frac {\mu^2}{\mu_c^2}  \qquad {\rm for} \qquad \mu<\mu_c.
\label{zthnf2}
\ee
This result is in agreement with the mean field result (\ref{phmftpq}). 
Indeed, the denominator of (\ref{zthpq}) is the product of the  Goldstone masses 
$\mu_c-\mu$, $\mu_c+\mu$, $\mu_c$ and $ \mu_c$, 
whereas the denominator of
(\ref{zthnf2}) is the product of four Goldstone masses 
$ \mu_c$. The free energy of both partition
functions is equal to $2nN_c\mu_c$.  

\begin{figure}[!t]
  \unitlength1.0cm
    \epsfig{file=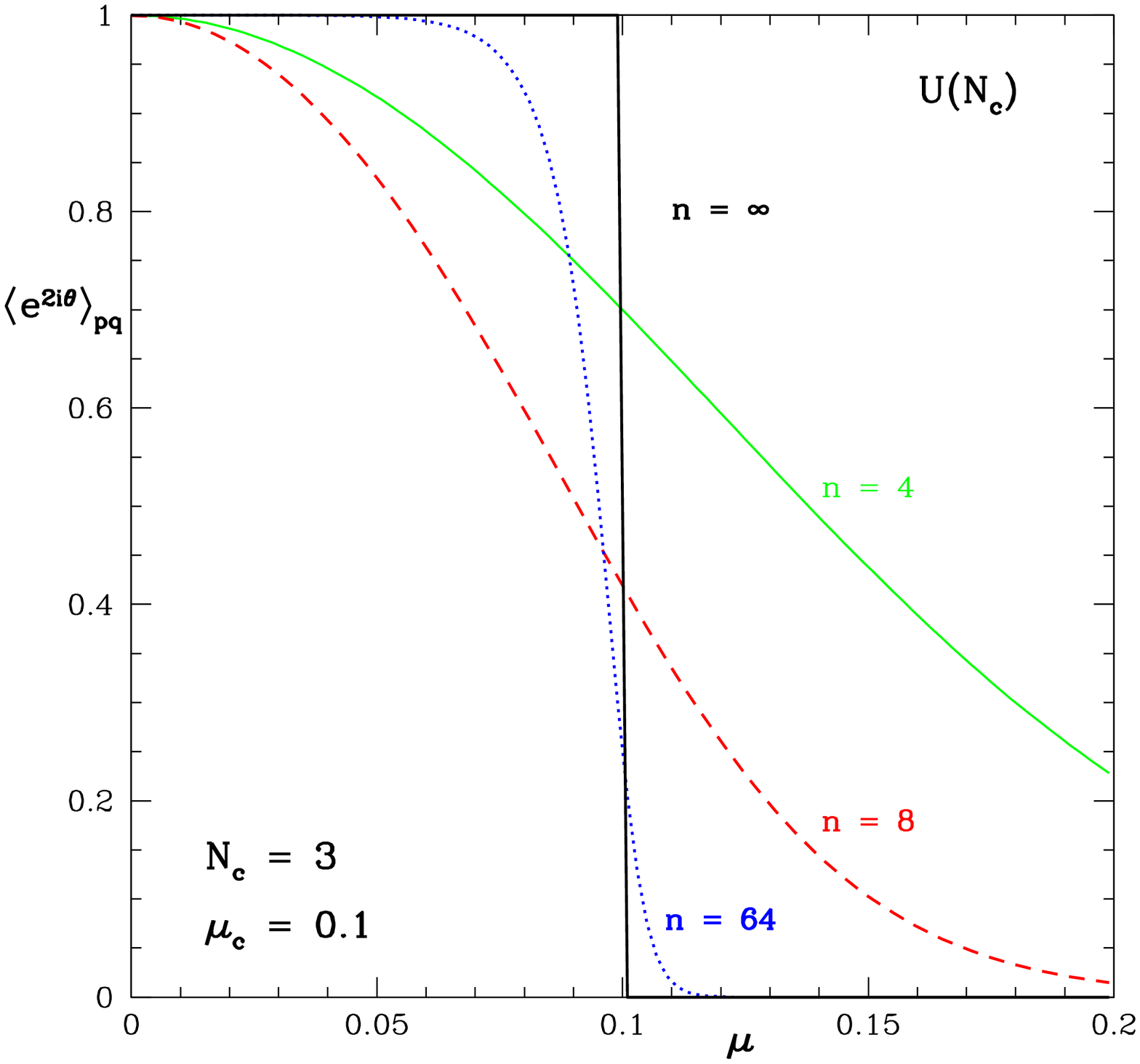,clip=,width=8cm}
    \epsfig{file=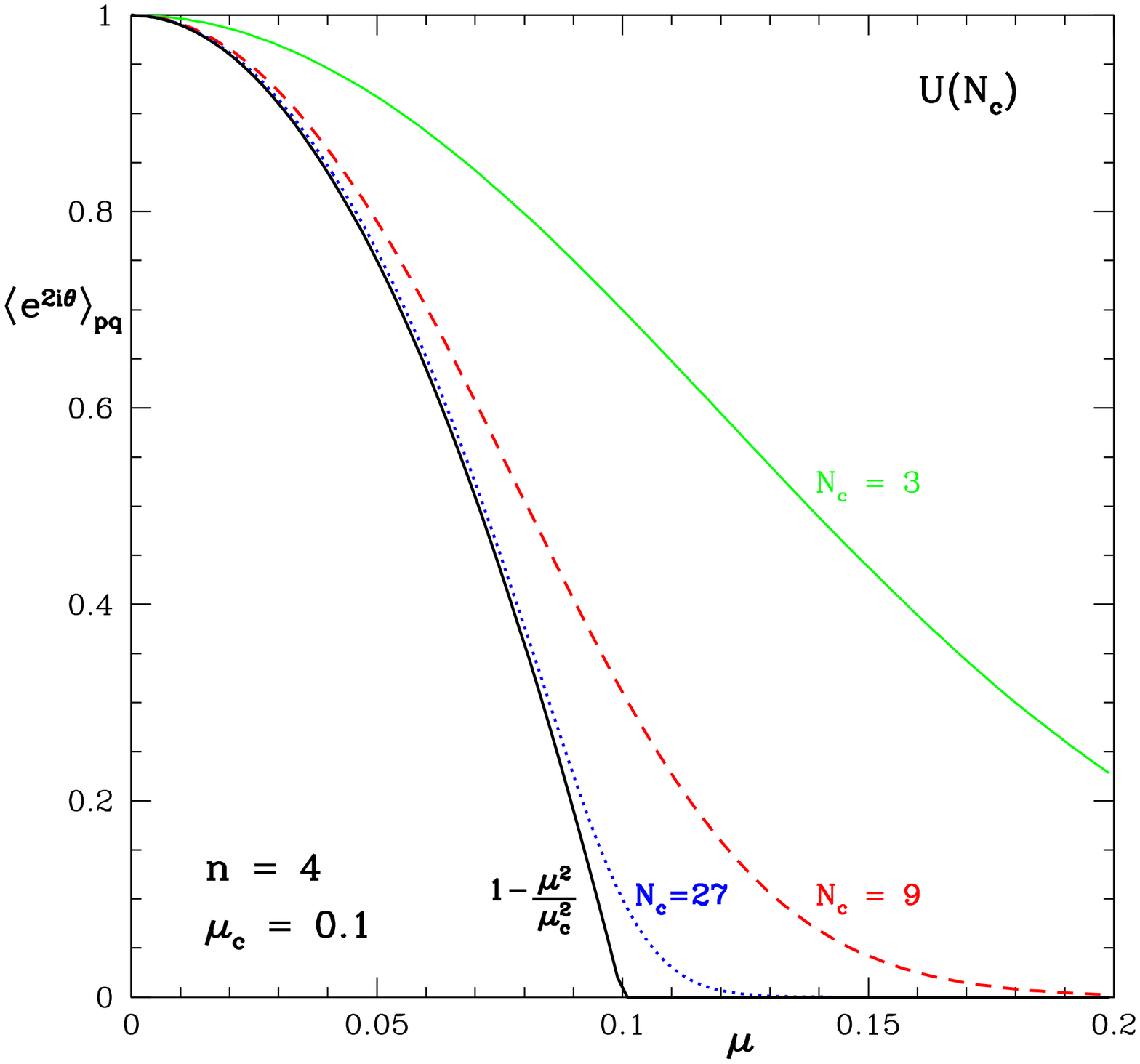,clip=,width=8cm}
  \caption{
The average phase factor for one-dimensional QCD
with gauge group $U(N_c)$
calculated as the ratio of the two flavor flavor partition function 
and the phase quenched partition function. In the left figure the number
of lattice points varies as indicated for fixed
$N_c=3$, and in the right figure
the number of colors varies as indicated for fixed $n=4$.}
 \label{fig:phase} 
\end{figure}

For $\mu > \mu_c$, the large $N_c$ limit of
the two flavor partition function is still given by
(\ref{znftlg}), but
the large $N_c \mu$ limit of the microscopic
 phase quenched partition function is now given by
\be
Z_{\rm pq} = \frac{e^{2nN_c \mu}}
{16n^4(\mu^2-\mu^2_c)\mu^2},
\ee
resulting in the average phase factor
\be
\langle e^{2i\theta} \rangle_{\rm pq} = \frac{Z_{N_f=2}}{Z_{\rm pq}} =
\frac{\mu^2}{\mu_c^2}
( \frac {\mu^2}{\mu_c^2} -1) e^{-2nN_c(\mu-\mu_c)} \qquad {\rm for} \qquad \mu>\mu_c\,  .
\ee
This result can also be obtained from the small $n\mu,n\mu_c$ expansion of
the large $N_c$ limit of the average phase factor (see Eq. (\ref{phlgnu})) at fixed 
$nN_c\mu_{c}$.
We conclude that  the sign problem becomes exponentially hard for $\mu>\mu_c$.

In Fig. \ref{fig:phase} we show the average phase factor for different values of $n$ and $N_c$
and a critical chemical potential of $\mu_c = 0.1$. In the thermodynamic limit at fixed $N_c$, the average
phase factor is one for $\mu < \mu_c$ and jumps to zero for $\mu>\mu_c$. In the right figure we observe
a rapid convergence to the mean field result discussed in section \ref{largen}.

\subsection{Average phase factor for full QCD}

The average phase factor with the $N_f$-flavor quark determinant as weight is defined by
\be
\langle e^{2i\theta} \rangle_{N_f}
&=& \frac 1 {Z_{N_f}(\mu_c,\mu)}\int dU \frac {\det D}{\det D^\dagger}{\det}^{N_f} D \nn \\
&=& \frac{e^{nN_fN_c\mu_c}}{Z_{N_f}(\mu_c,\mu)}\int dU \frac{\det (1-U e^{n\mu-n\mu_c})
∑\det(1- U^\dagger e^{-n\mu-n\mu_c})}
{\det (1-U e^{-n\mu-n\mu_c})\det(1- U^\dagger e^{n\mu-n\mu_c})}
{\det}^{N_f} (1-U e^{n\mu-n\mu_c})
{\det}^{N_f}(1- U^\dagger e^{-n\mu-n\mu_c}).
\nn \label{phfull}
\ee
We evaluate this integral exactly for $N_f=0$ and $N_f=1$ but 
 only give its
large $N_c$ limit
for other values of $N_f$.

Using the CFZ-formula (\ref{cfz}) \cite{CFZ}  we obtain in the quenched case 
\be 
\langle e^{2i\theta} \rangle_{N_f=0} 
=\frac{(1-e^{-2n(\mu+\mu_c)})(1-e^{2n(\mu-\mu_c)})}{(1-e^{-2n\mu_c})^2}
+ e^{-2nN_c \mu_c} \frac{(1-e^{-2n\mu})(1-e^{2n\mu})}
{(1-e^{2n\mu_c})(1-e^{-2n\mu_c})} \quad {\rm for} \quad \mu<\mu_c\, .
\label{phquen}
\ee
The large $N_c$ limit of this result  coincides with the large $N_c$ limit of
the phase quenched average phase factor.
In the microscopic
limit where $\mu_cn N_c$ and $\mu nN_c$ remain fixed for $N_c \to \infty$ we
obtain
\be
\langle e^{2i\theta} \rangle_{N_f=0} = 1 -  \frac {\mu^2} 
{\mu_c^2} -\frac{\mu^2}{\mu_c^2}e^{-2nN_c\mu_c}.
\label{phmicro3}
\ee
In the large $n N_c\mu_c$ limit the second term does not contribute resulting in 
\be
\langle e^{2i\theta} \rangle_{N_f=0} = 1 -  \frac {\mu^2} 
{\mu_c^2},
\ee
in agreement with the results obtained in \cite{SVphase1} 
(see subsection \ref{largen}). In the thermodynamic limit at fixed $N_c$ the average
phase factor converges to one for $\mu < \mu_c$.
\begin{figure}[t!]
  \unitlength1.0cm
    \epsfig{file=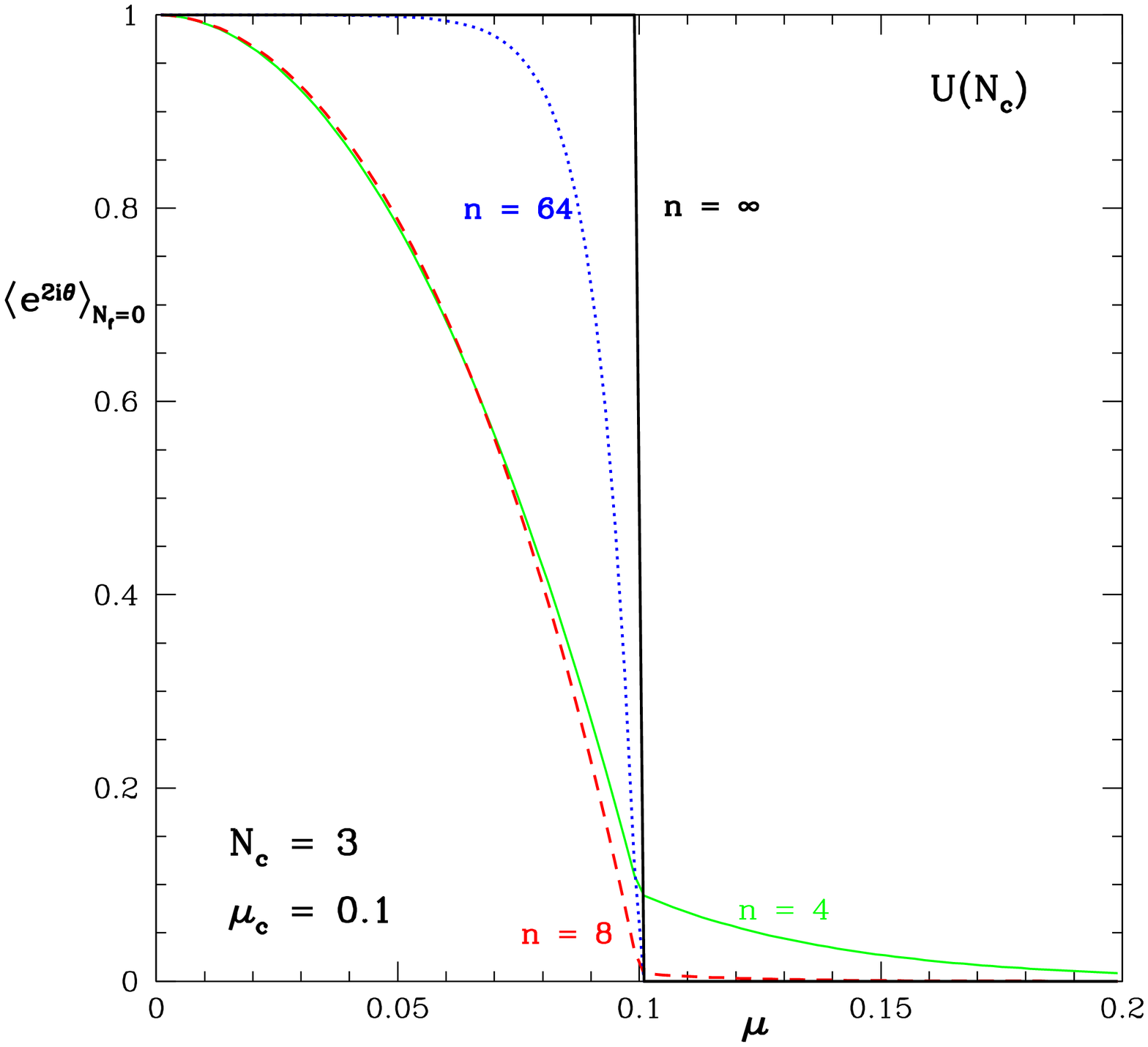,clip=,width=8cm}
    \epsfig{file=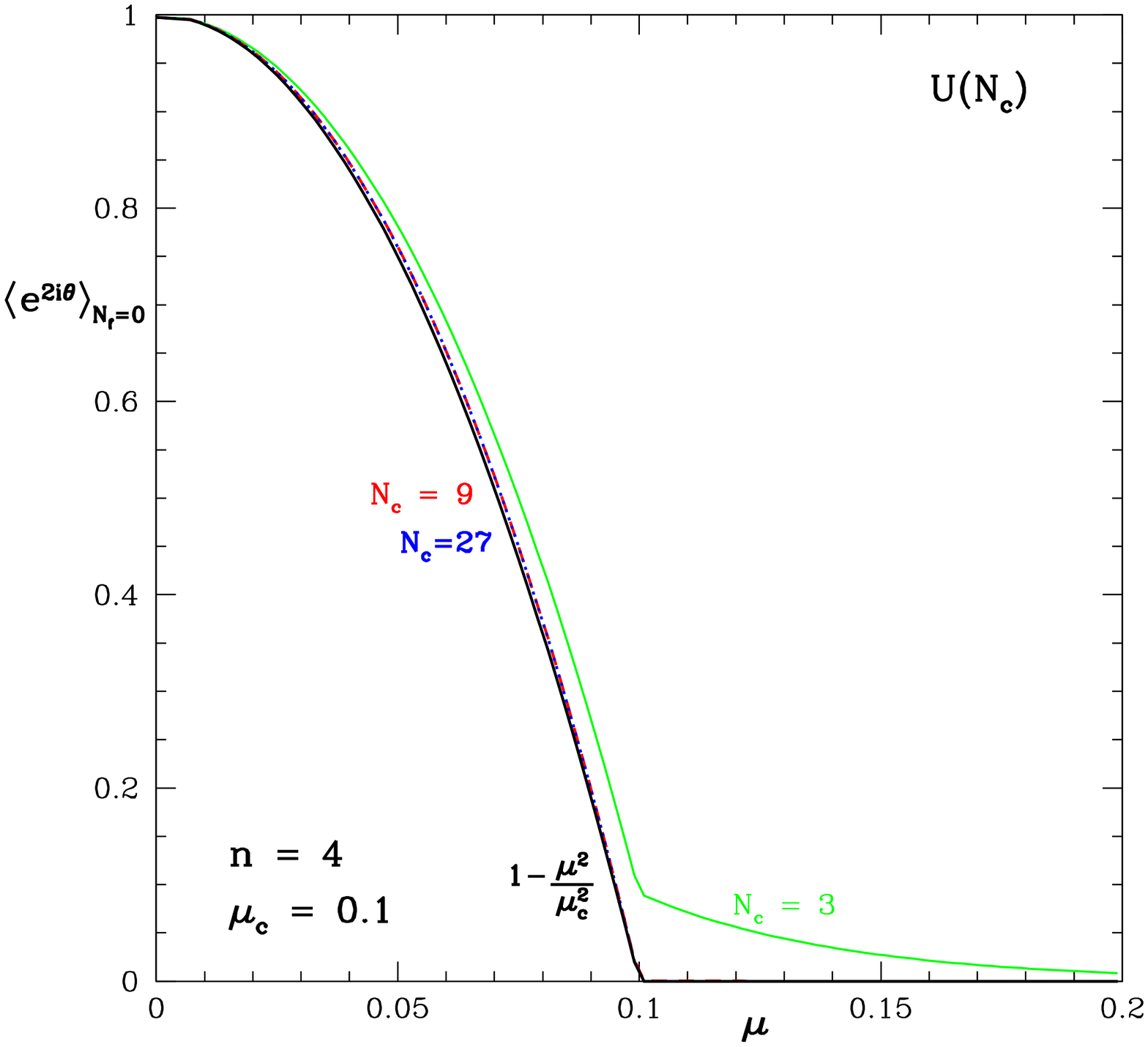,clip=,width=8cm}
  \caption{
The quenched average phase factor for one-dimensional QCD
with gauge group $U(N_c)$. In the left figure $N_c=3$ and
the number of lattice points is as indicated. In the right figure the
number of colors varies whereas $n=4$.}
  \label{fig2} 
\end{figure}

After rewriting the determinants in (\ref{phfull})  as
\be
\frac
{\det(1- U^\dagger e^{-n\mu-n\mu_c})}
{\det(1- U^\dagger e^{n\mu-n\mu_c})}
\to
e^{-2nN_c\mu}\frac{\det(1-  Ue^{n\mu+n\mu_c})}
{\det(1- U e^{-n\mu+n\mu_c})},
\label{rewrite}
\ee
    the denominator of (\ref{phfull}) 
can be expanded in powers of $U$ for $\mu >\mu_c$.  
For $N_f =0$ only the constant term in the integrand yields a non-vanishing
result.
We thus find
\be
\langle e^{2i\theta} \rangle_{N_f=0}  =e^{-2nN_c \mu}
\qquad {\rm for} \qquad 
\mu>\mu_c,
\ee
so that the average phase factor vanishes in the large $nN_c\mu$ limit.

In Fig. \ref{fig2} we show the quenched average phase factor for various values of $n$ and $N_c$.
Also in this case the average phase factor jumps from 1 to 0 at $\mu =\mu_c$ in the thermodynamic limit.
The convergence to the mean field result for increasing $N_c$ (right figure) is very rapid.
This can be understood from the expansion of the microscopic result
given by $1-(\mu^2/\mu_c^2)(1-\frac 13(n\mu)^2 +\cdots$.

Finally we  calculate the average phase factor for an arbitrary
number of flavors in the large $N_c$ limit.
The CFZ formula can again be applied after inserting  the factor
\be
{\det}^{N_f} (1-U e^{-\alpha})
{\det}^{N_f}(1- U^\dagger e^{-\beta})\nn
\label{insert}
\ee
in the denominator and taking the limit $\alpha \to \infty $ 
and $\beta \to \infty$ at the end of the calculation.

For $\mu < \mu_c$, 
 the leading
order large $N_c$ result is given by 
the identity permutation in the sum over permutations in the CFZ formula.
This results in
\be
\langle e^{2i\theta} \rangle_{N_f} &=&
\left (\frac {(1-e^{-2n(\mu+\mu_c)})(1-e^{2n(\mu-\mu_c)})}
{(1-e^{-2n\mu_c})^2}\right )^{N_f+1} \quad {\rm for} \quad \mu < \mu_c .
\label{zphq}
\ee
The microscopic limit of this result is given by
\be
\langle e^{2i\theta} \rangle_{N_f} =
\left [1 -  \frac {\mu^2}
{\mu_c^2}  \right ]^{N_f+1} \quad {\rm for} \quad \mu < \mu_c,
\ee
in agreement with the discussion given in section  \ref{largen}.

For $\mu> \mu_c$ the CFZ formula can again be applied after rewriting
the partition function according to (\ref{rewrite}) and the insertion (\ref{insert}).
Because of degenerate prefactors, all critical chemical potentials that
occur in the combination $\mu+\mu_c$ have to be taken different. 
After carefully taking limits we find
\be
Z_{N_f+1|1^*}(\mu_c,\mu)= \frac{N_c^{N_f}}{N_f!} e^{ - 2 \mu n N_c}
\frac{e^{\mu_c n N_c N_f}} {(1-e^{-2\mu_c n})^{N_f^2}}
\frac{(1-e^{-2n(\mu+\mu_c)})^{N_f} (1-e^{-2n \mu})^{N_f} }
{(1-e^{-2n\mu_c})^{N_f}}.
 \ee
The average phase factor  given by
\be
\langle e^{2i\theta}\rangle_{N_f} = \frac{Z_{N_f+1|1^*}}{Z_{N_f}}\sim e^{-2nN_c\mu}
\quad {\rm for} \quad \mu > \mu_c
\ee
 vanishes in the thermodynamic limit. The exact result becomes rather cumbersome
even for small values of $N_f$. As an illustration we give the result for
$Z_{2|1^*}(\mu_c,\mu)$ in Appendix A .

\section{Average Phase Factor for $SU(N_c)$}

To calculate integrals over $SU(N_c)$ we use the identity
\be
\int_{SU(N_c)} dU \cdots 
= \sum_{p=-\infty}^\infty \int_{U(N_c)} d U {\det}^p(U) \cdots \, .
\ee
For the partition functions discussed below, with a few exceptions
the sum truncates to a small number of terms.

\subsection{$SU(N_c)$ QCD Partition Function }

Let us first calculate the $1d$ $SU(N_c)$ QCD partition function for $N_f=1$. 
The partition function is defined by
\be
Z_{N_f}^{SU(N_c)}(\mu_c,\mu)=\int_{SU(N_c)} dU {\det}^{N_f} D
=       \sum_{p=-\infty}^\infty       \int_{U(N_c)} dU {\det}^p U e^{n N_c \mu_c}
  {\det}^{N_f} (1+U e^{n\mu-n\mu_c})
{\det}^{N_f}(1+ U^\dagger e^{-n\mu-n\mu_c}).\hspace*{0.5cm}
\ee
For $p \ge N_f$ the integrand can be rewritten in terms of  determinant
of matrices $U$ only 
\be
Z_{N_f,p}(\mu_c,\mu)  &=& \int_{U(N_c)} dU {\det}^p U {\det}^{N_f}  D \nn\\
&=&              \int_{U(N_c)} dU 
e^{n N_c N_f \mu_c}  {\det}^{p-N_f}(U) {\det}^{N_f} (1+U e^{n\mu-n\mu_c})
{\det}^{N_f}(U+ e^{-n\mu-n\mu_c})
\, .
\ee
For $p> N_f$ it follows immediately that the integral vanishes. 
For $p =N_f$ only the constant term inside the determinants contributes to
integral resulting in 
\be
Z_{N_f,p=N_f}(\mu_c,\mu) &=&  e^{-nN_fN_c \mu},\nn\\
Z_{N_f,p> N_f}(\mu_c,\mu) &=&  0.
\label{fullsim}
\ee
For $p<0$ we combine ${\det}^p U = {\det}^{-p} U^\dagger$ with the factor
${\det}^{N_f} (1-U e^{n\mu-n\mu_c})$. We then find 
\be
Z_{N_f,p=-N_f}(\mu_c,\mu) &=&  e^{nN_fN_c \mu},\nn  \\
Z_{N_f,p< -N_f}(\mu_c,\mu) &=&  0.
\label{suone}
\ee
Using that the hermitean conjugate of a unitary matrix is also unitary we
obtain
\be
Z_{N_f,-p}(\mu_c,\mu  ) = Z_{N_f,p}( \mu_c,-\mu).
\label{symsu}
\ee
For $|p| < N_f$ the integral can be calculated by means of the CFZ formula
by choosing all $\mu_c$ different and introducing the limit
\be 
\det U =\lim_{\alpha \to \infty} \det( U - e^{-\alpha}).\label{dlim}
\ee
The limit of degenerate $\mu_c$
and $\alpha \to \infty$ is taken at the end of the calculation. 
We will only give  exact results for $N_f=1$ and $N_f =2$.

Using (\ref{fullsim},\ref{suone}) and the result for $U(N_c)$ we obtain for $N_f=1$ 
\be
Z_{N_f = 1}^{SU(N_c)}(\mu_c,\mu) = e^{-nN_fN_f \mu} +e^{nN_fN_f \mu} +
\frac{ \sinh(n (N_c+1) \mu_c)}{\sinh(n\mu_c)},
\ee
in agreement with earlier work by Bilic and Demeterfi \cite{bilic}.

For $N_f =2$, the only integrals that have not yet been calculated  are 
those for  $p =  \pm 1$ which are related
by (\ref{symsu}). For the $p=1$ contribution we find
\be
Z_{N_f=2,p=1}(\mu_c,\mu) = e^{nN_c(\mu_c -\mu)}\left [N_c
\frac{e^{-2n(N_c+1)\mu_c} +e^{2n\mu_c}}{(e^{n\mu_c}- e^{-n\mu_c})^2}
-\frac{e^{-2nN_c\mu_c}(e^{-3n\mu_c} -3e^{-n\mu_c})
 - e^{ 3n\mu_c} +3e^{ n\mu_c}   }{(e^{n\mu_c}- e^{-n\mu_c})^3}\right ] .
\ee

For $\mu< \mu_c$ the large $nN_c\mu$ limit of the $N_f=2$ partition function
is dominated by the $p=0$ term, whereas for $\mu>\mu_c$ the large $nN_c \mu_c$ limit
is given  by the $p=-2$ term. 
For $\mu = \mu_c$ the large $N_c$  limit is dominated by the $p=-1$ with $1/N_c$ corrections
from the $p=0$ and $p=-2$ terms.
We thus find as
 leading large $N_c$ result
\be
Z_{N_f=2}^{SU(N_c)}(\mu_c,\mu)  &=& \frac {e^{2n(N_c +2)\mu_c}}{16\sinh^4(n\mu_c) } \quad {\rm for} \quad \mu <\mu_c,\nn\\
Z_{N_f=2}^{SU(N_c)}(\mu_c,\mu)  &=& e^{2nN_c \mu} \quad {\rm for} \quad \mu> \mu_c, \nn \\
Z_{N_f=2}^{SU(N_c)}(\mu_c,\mu)  &=& \frac{N_c e^{2n(N_c+1)\mu_c}
}{4\sinh^2(n\mu_c)} \left[ 1 + O(N_c^{-1})\right ]
\quad {\rm for} \quad \mu= \mu_c .
\ee
Both for $\mu< \mu_c$ and $\mu>\mu_c$
 the terms that are canceled by the integration over the unitary group
are subleading in the thermodynamic limit.
In other words there is no serious sign problem.

By inspection one can easily show that the dominance of the $p=0$ terms for $\mu<\mu_c$ and
the $p=-N_f$ term for $\mu >\mu_c$ is a feature of the large $N_c$ limit that is valid for
any number of flavors. Therefore,
for $\mu<\mu_c$, the chiral condensate is the
same as for the $U(N_c)$ theory. For $ \mu > \mu_c$ though, the 
 dominant  $p=-N_f$ term is mass-independent.
This results in a vanishing chiral condensate so that 
the large $N_c$ limit of the $U(N_c)$ theory and the $SU(N_c)$ theory is 
different.

\subsection{The Phase Quenched Partition Function for $SU(N_c)$}

A second ingredient for the average phase factor is the 
phase quenched $SU(N_c)$ partition function. Using the same arguments
as for the full QCD partition function, one easily obtains
\be
Z_{1+1^*,|p|>2}(\mu_c,\mu ) &=& 0,\nn \\
Z_{1+1^*,|p|=2}(\mu_c,\mu ) &=& 1, \nn \\
Z_{1+1^*,-p}(\mu_c,\mu )&= &Z_{1+1^*,p}(\mu_c,-\mu).  
\ee
The partition function for $p=0$ is the $U(N_c)$ partition function which
was already given in (\ref{zpq}). What  remains to  be
calculated is  the partition function for $p=+1$. Using the CFZ formula one
easily arrives at
\be
Z_{1+1^*,p=1}(\mu_c,\mu )&=&\frac {1}{4\sinh(n\mu)\sinh(n\mu_c)}
\left(\frac{\sinh(n(N_c+2)(\mu+\mu_c))}{\sinh(n(\mu+\mu_c))} 
-\frac{\sinh(n(N_c+2)(\mu-\mu_c))}{\sinh(n(\mu-\mu_c)))}\right )
 .
\ee

Both for $\mu<\mu_c$ and $\mu>\mu_c$,
the leading large $N_c$ result
of the phase quenched partition function resides in the $p=0$ term given in
Eq. (\ref{zpqlnc}).
\begin{figure}[t!]
  \unitlength1.0cm
    \epsfig{file=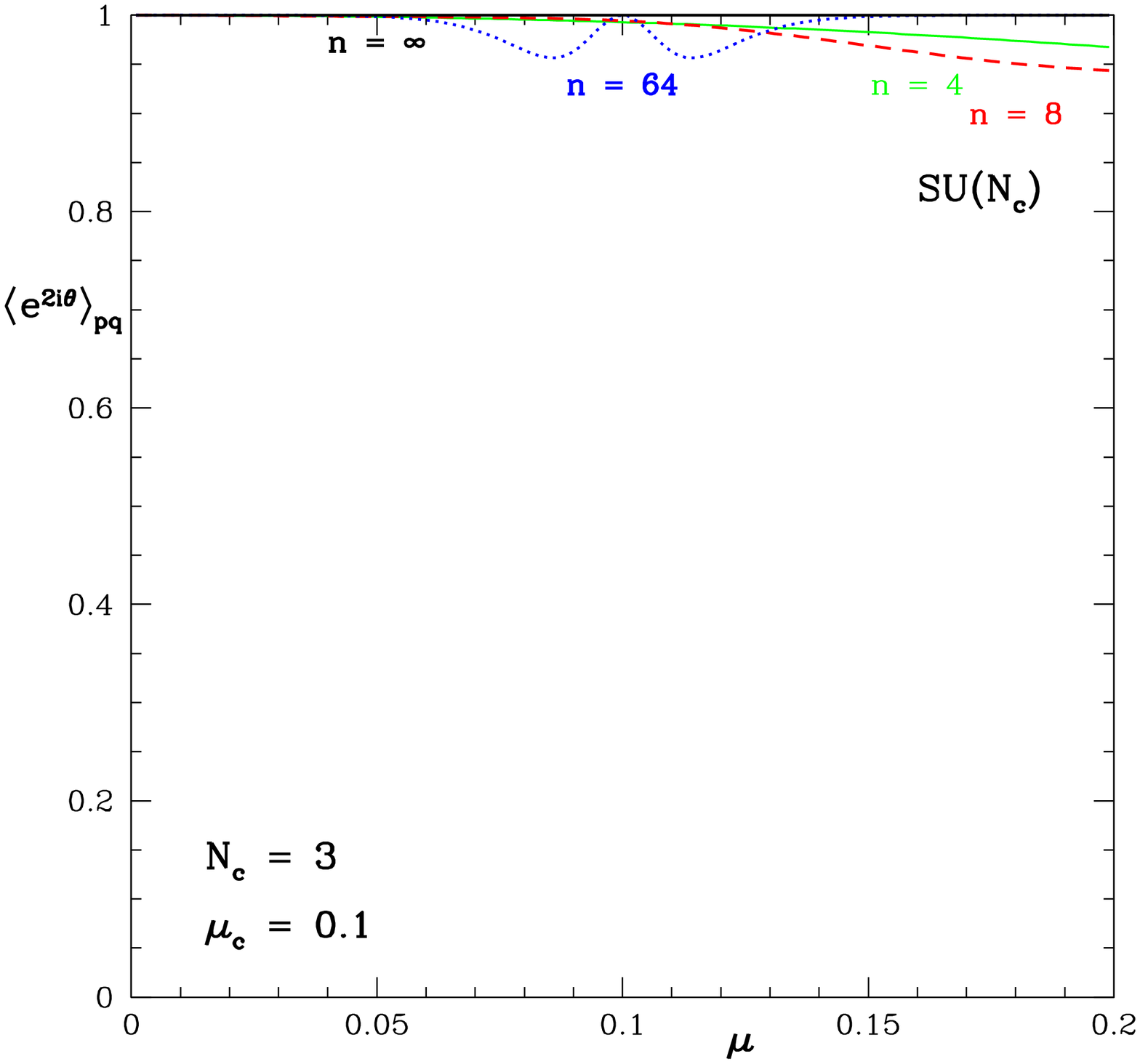,clip=,width=8cm}
    \epsfig{file=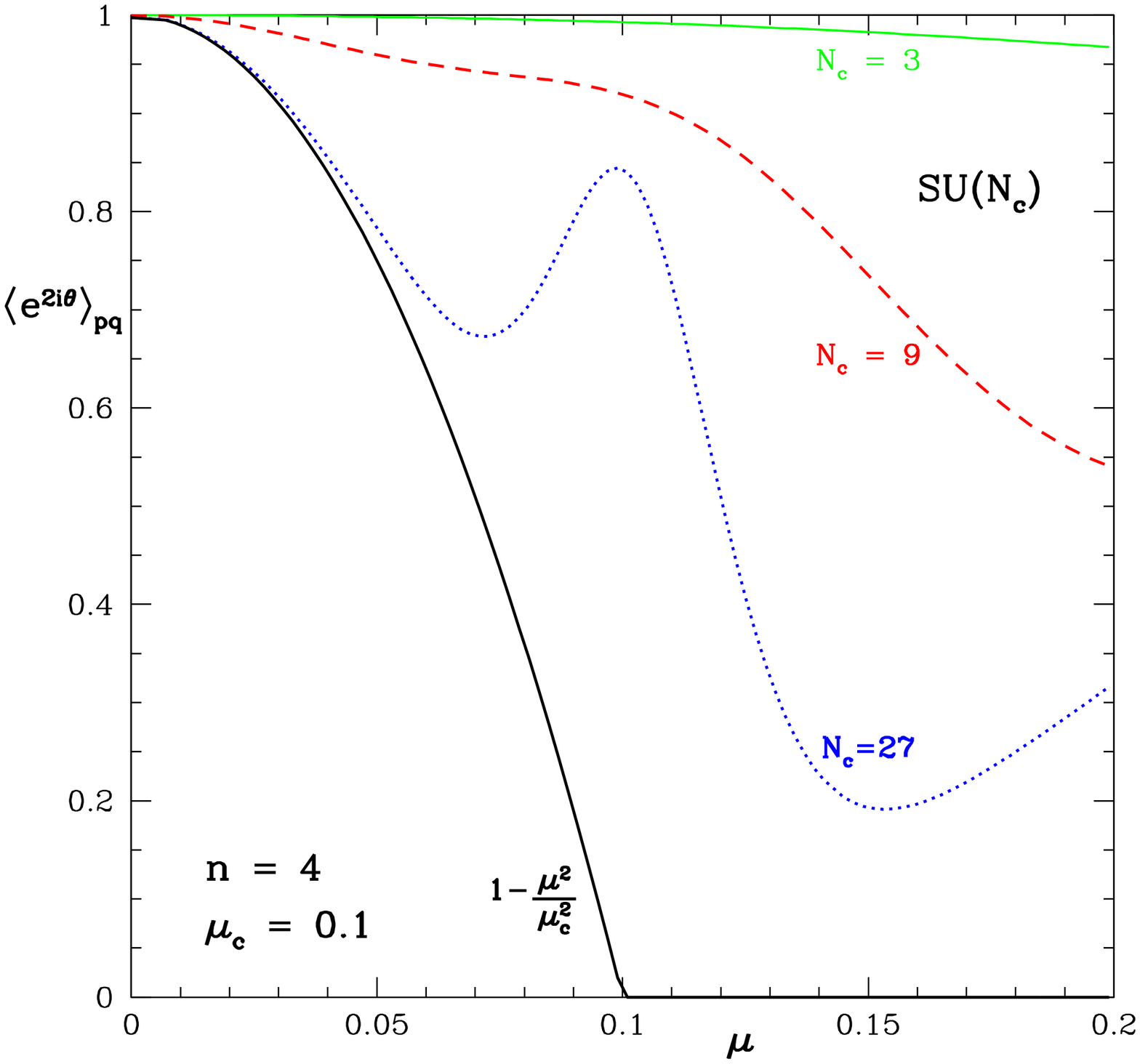,clip=,width=8cm}
  \caption{
The phase quenched average phase factor for one-dimensional QCD
with gauge group $SU(N_c)$. In the left figure $N_c=3$ and
the number of lattice points is as indicated. In the right figure the
number of colors varies whereas $n=4$.}
  \label{fig3} 
\end{figure}

\subsection{Average Phase Factor for $SU(N_c)$ }

The average phase factor can be either obtained from the ratio of the 
full QCD partition function and the phase quenched partition function or
can be calculated with the full QCD partition function as weight 
(see Eqs. (\ref{eq2}) and (\ref{eq3})).
In the first case we find for large $N_c$,
\be
 \langle e^{2i\theta} \rangle_{\rm pq}^{SU(N_c)}
=\left \{ \begin{array}{ccc}
\frac{\sinh(n(\mu_c-\mu)) \sinh(n(\mu_c+\mu))}{\sinh^2(n\mu_c)}
 &\qquad {\rm for} \qquad & \mu <\mu_c, \\ && \\
16\sinh(n(\mu-\mu_c)) \sinh(n(\mu_c+\mu))\sinh^2(n\mu)
 &\qquad {\rm for} & \mu> \mu_c,\\ && \\ 
1- e^{-4n\mu_c} &\qquad {\rm for} & \mu= \mu_c  .
\end{array} \right . 
\label{phlgn}
\ee
The microscopic limit of the average phase factor is obtained by expanding this result
for small $\mu_c,\mu$. We again find the usual mean field result
\be
\langle e^{2i\theta} \rangle_{\rm pq}^{SU(N_c)} = \left ( 1 - \frac{\mu^2}{\mu_c^2} \right ) 
\theta(\mu_c-\mu).
\ee
In the thermodynamic limit, 
the average phase factor converges to one in both for $\mu<\mu_c$ and
$\mu>\mu_c$.

Exact results for the average phase factor are displayed in  Fig. \ref{fig3}. 
At 
finite $N_c$,  we observe
a rapid convergence to the asymptotic value of 1. 
At fixed $n$, the approach to the large $N_c$ limit
is slow. Also clearly visible is that  for $\mu \approx \mu_c$ the 
corrections to the microscopic limit are large.
As was already discussed before, this is due
to the $p=-1$ contribution.

In the quenched case the average phase factor for $SU(N_c)$ defined 
according to Eq. (\ref{eq3}) 
can be written as
\be
 \langle e^{2i\theta} \rangle_{N_f=0}^{SU(N_c)} 
&=& \sum_{p =-\infty}^\infty 
\int_{U\in U(N_c)} dU {\det}^p(U) 
\frac{\det (1+U e^{n\mu-n\mu_c})
\det(1+ U^\dagger e^{-n\mu-n\mu_c})}
{\det (1+U e^{-n\mu-n\mu_c})\det(1+U^\dagger e^{n\mu-n\mu_c})}.
\label{phasdef}
\ee 
The integrals can again be calculated by means of the CFZ formula.
For $N_c\ge 3$  one can easily show that the limit introduced in (\ref{dlim})
gives vanishing results for  $|p|> 2$. However, for $N_c=1$ and $N_c=2$
additional terms may contribute to this limit.
For $N_c=2$, where the average
phase factor is equal to one, contributions for $|p|>2$ vanish for 
$\mu<\mu_c$, but terms with $p<-2$ are nonzero for $\mu>\mu_c$. The formulae
for $|p|\le 2$ given below are valid for $N_c=2$.  
For $N_c=1$ the integrals in (\ref{phasdef}) only vanish for
$p \ge 2$ and $\mu>\mu_c$. Also the general answer for $p=-2$ is not correct in
this case.

For $p=- 1$ we obtain
\be
 \langle e^{2i\theta} \rangle_{N_f=0}
=\left \{ \begin{array}{ccc}
e^{-N_c n(\mu_c -\mu)} \frac{(1-e^{-2n(\mu_c +\mu)})(1-e^{-2n\mu})}
{1-e^{-2n\mu_c}}& \qquad & {\rm for} \qquad\mu <\mu_c , \\ & & \\
\frac{e^{nN_c(\mu_c-\mu)}(1-e^{-2n\mu})}{1-e^{-2n\mu_c}}
[(1-e^{-2 n \mu_c-2n\mu})-e^{-2nN_c\mu_c}(e^{-2n\mu_c } - e^{-2n\mu})]
& \qquad & {\rm for} \qquad \mu>\mu_c  . \end{array}\right .
\ee
For $p=1$ we find
\be
 \langle e^{2i\theta} \rangle_{N_f=0}
=\left \{ \begin{array}{ccc}
e^{-N_cn(\mu_c +\mu)} \frac{(1-e^{-2n(\mu_c -\mu)})(1-e^{2n\mu})}
{1-e^{-2n\mu_c}} &\qquad {\rm for} \qquad & \mu <\mu_c, \\ && \\
0 &\qquad {\rm for} & \mu> \mu_c  . \end{array}\right .
\ee
Finally, for    $p=-2$ the result is
\be
 \langle e^{2i\theta} \rangle_{N_f=0}
=\left \{ \begin{array}{ccc}
 0  &\qquad {\rm for} \qquad & \mu <\mu_c,\\ && \\
(1-e^{-2n\mu})(1-e^{-2n(\mu+\mu_c)})(1-e^{2n(\mu_c-\mu)})(1-e^{-2n\mu})
&\qquad {\rm for} & \mu> \mu_c .\end{array}\right .
\ee
For $p=2$ the average phase factor vanishes.
Notice that the relation
\be
 \langle e^{2i\theta} \rangle_{N_f=0,-p}(\mu_c,\mu) = 
 \langle e^{2i\theta} \rangle_{N_f=0,p}(\mu_c,-\mu) 
\ee
is also valid in this case. It can be used to derive results for negative values of $\mu$.
\begin{figure}[t!]
  \unitlength1.0cm
    \epsfig{file=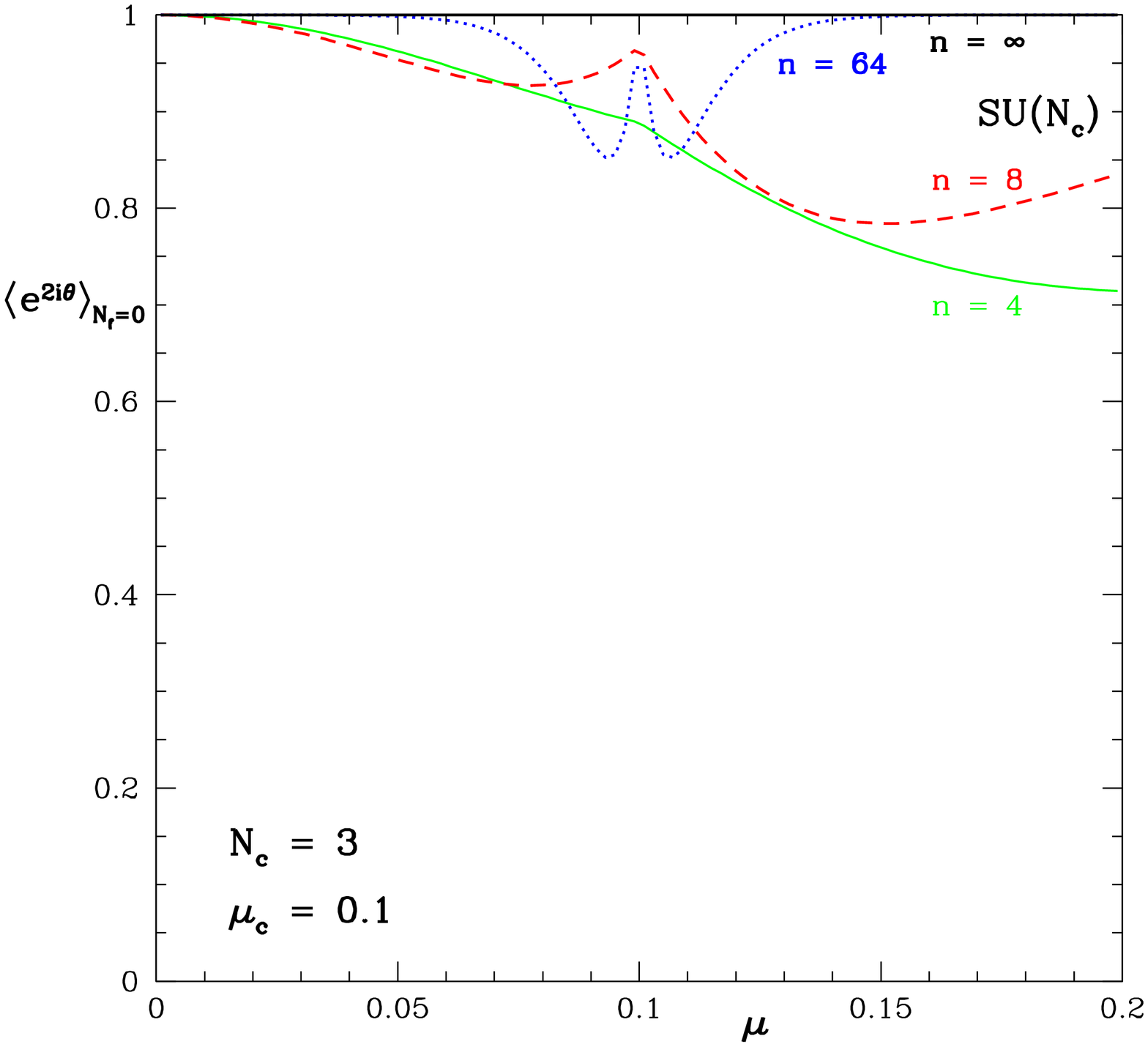,clip=,width=8cm}
    \epsfig{file=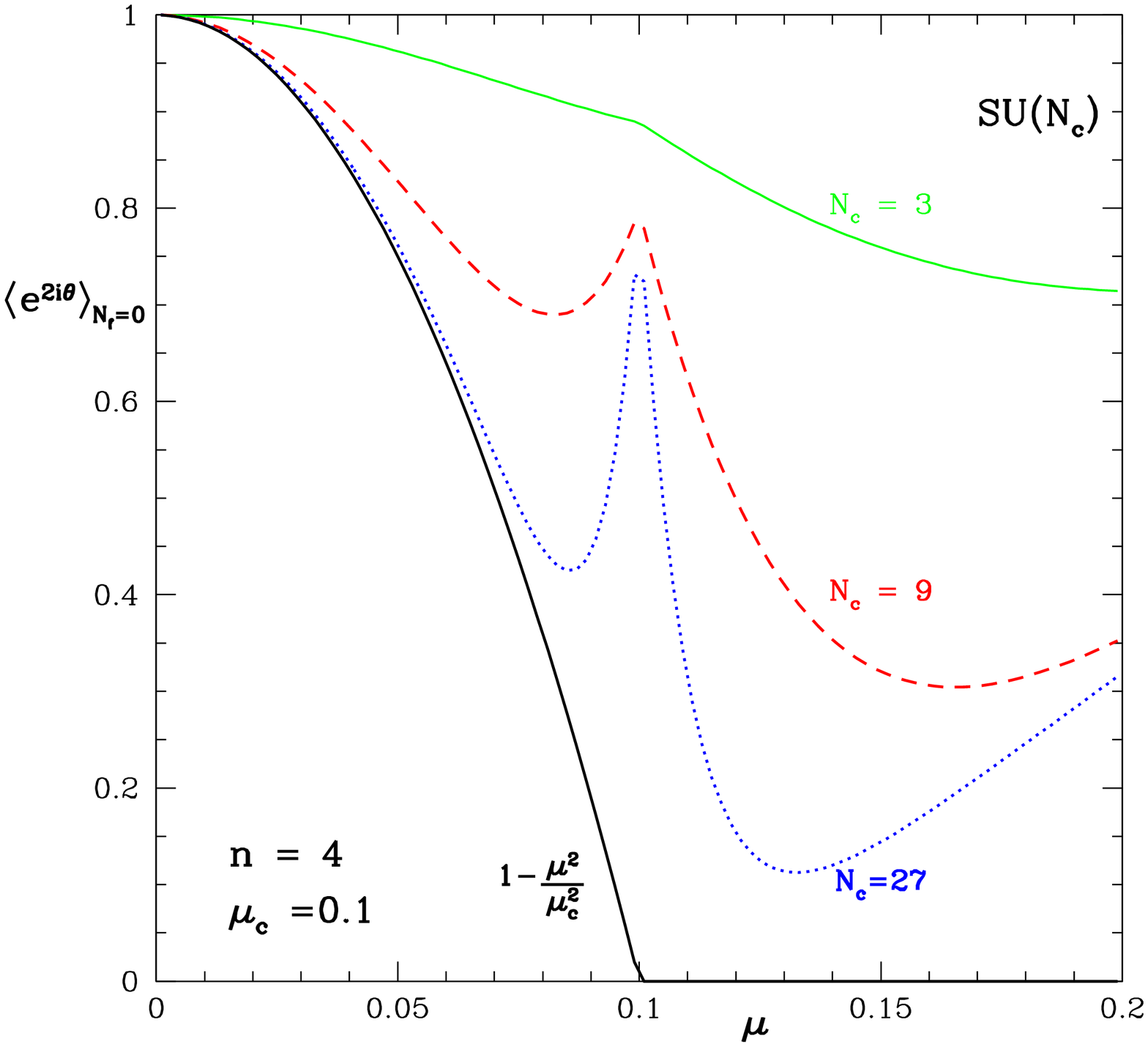,clip=,width=8cm}
  \caption{
Average phase factor for one-dimensional QCD
with gauge group $SU(N_c)$. In the left figure $N_c=3$ and
the number of lattice points is as indicated. In the right figure the
number of colors varies whereas $n=4$.}
  \label{fig4} 
\end{figure}

In the large $N_c$ limit, the average phase factor  
is dominated by the $p=0$ contribution for $\mu<\mu_c$
and by the $p=-2$ contribution for $\mu>\mu_c$. For $\mu = \mu_c$, the
$p=0$, $p=-1$ and $p=-2$ contributions are of equal order in $N_c$, but only the $p=-1$ term is nonvanishing. 
As large $N_c$ limit
of the average phase factor we thus find
\be
 \langle e^{2i\theta} \rangle_{N_f=0}^{SU(N_c)} 
=\left \{ \begin{array}{ccc}
\frac{\sinh(n(\mu_c-\mu)) \sinh(n(\mu_c+\mu))}{\sinh^2(n\mu_c)}
 &\qquad {\rm for} \qquad & \mu <\mu_c, \\ && \\
16\sinh(n(\mu-\mu_c)) \sinh(n(\mu_c+\mu))\sinh^2(n\mu)
 &\qquad {\rm for} & \mu> \mu_c,\\ && \\ 
1- e^{-4n\mu_c}&\qquad {\rm for} & \mu= \mu_c , 
\end{array} \right . 
\ee
which is the same expression as obtained in (\ref{phlgn}) for the phase quenched
average phase factor.
The microscopic limit of the average phase factor is again given by the small
$\mu$-$\mu_c$ expansion of this result which is equal to 
 the usual mean field result
\be
 \langle e^{2i\theta} \rangle_{N_f=0}^{SU(N_c)} = \left(1-\frac {\mu^2}{\mu_c^2} \right )
\theta(\mu_c-\mu).
\ee

In Fig. 4 we show the exact result for the $SU(N_c)$ average phase factor.
We observe a rapid approach to the thermodynamic limit at fixed $N_c$ just
like in Fig. 3.

In the large $N_c$ limit the result for the quenched average phase calculated according 
to (\ref{phasdef}) is also given by (\ref{phlgn}).
As already was argued in the introduction, we conclude that QCD in 
one dimension  with $SU(N_c)$ as gauge
group does not have a serious sign problem, not only  for $\mu < \mu_c$, 
but also for
$\mu > \mu_c$. Indeed one-dimensional QCD at nonzero chemical potential could be 
simulated reliably by the Glasgow method \cite{maria1d}.

\section{The Sign Problem at Nonzero Temperature}

\begin{figure}[!t]
\begin{center}
  \unitlength1.0cm  \hspace*{-1cm}
    \epsfig{file=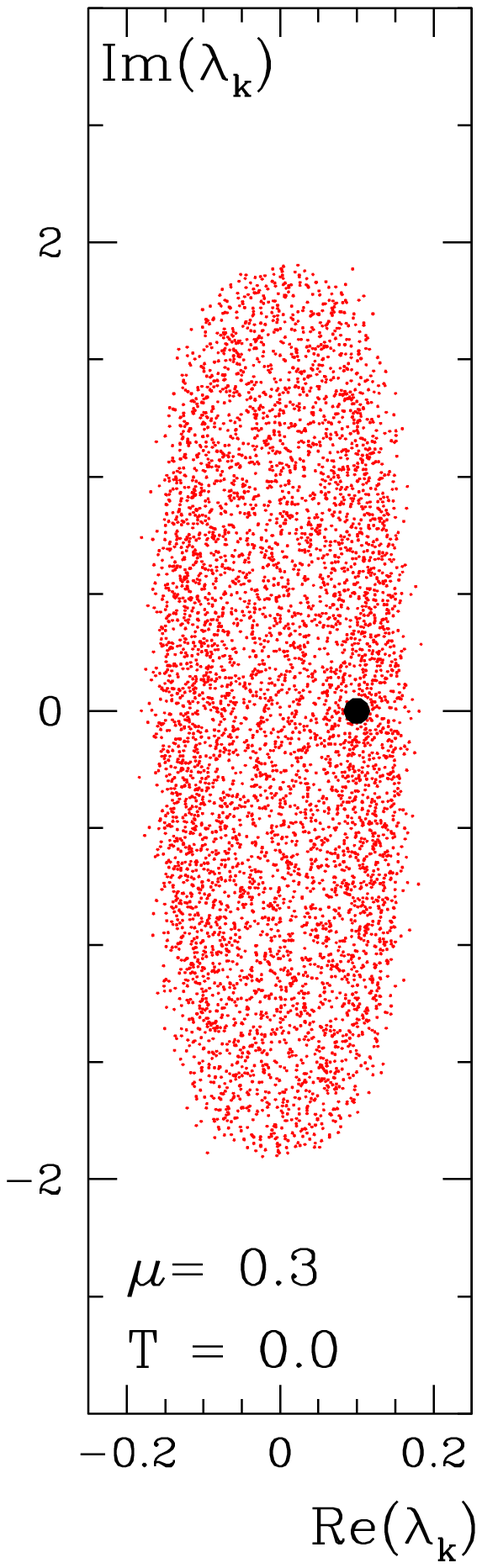,clip=,width=10cm}\hspace*{-6.5cm}
    \epsfig{file=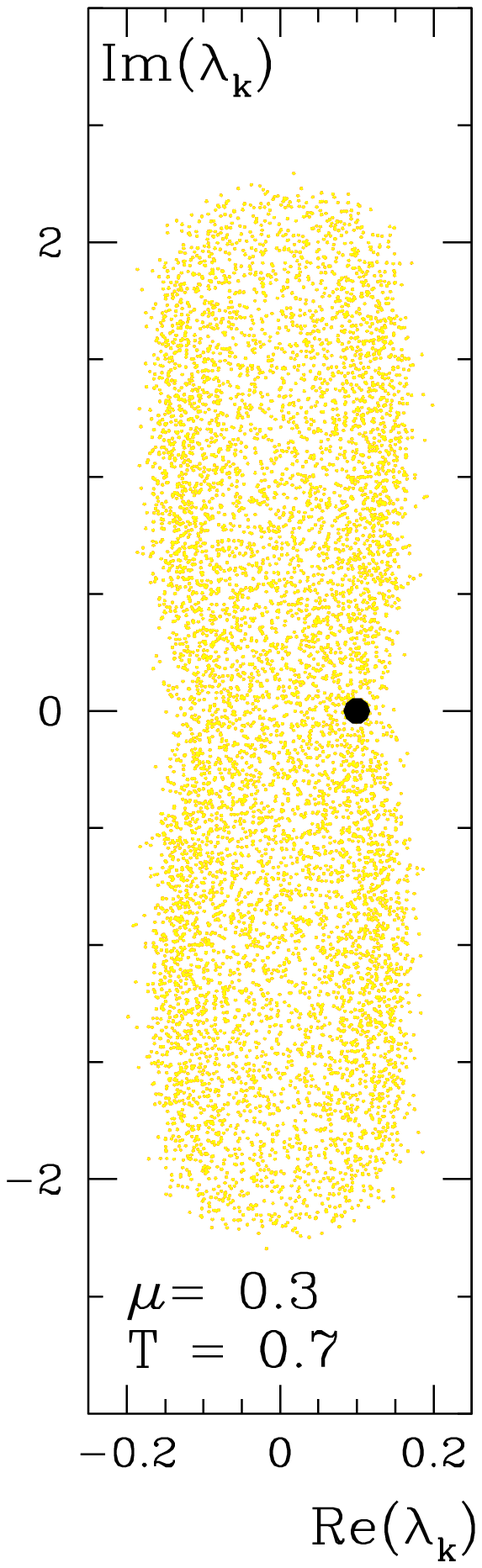,clip=,width=10cm}\hspace*{-6.5cm}
    \epsfig{file=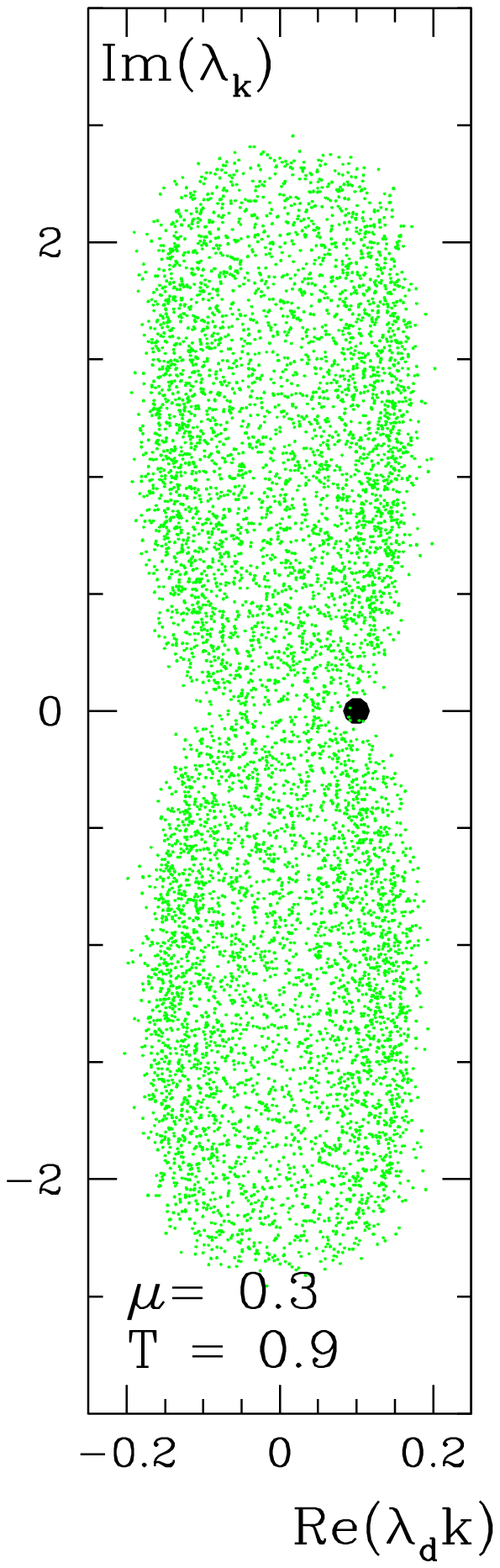,clip=,width=10cm}\hspace*{-6.5cm}
    \epsfig{file=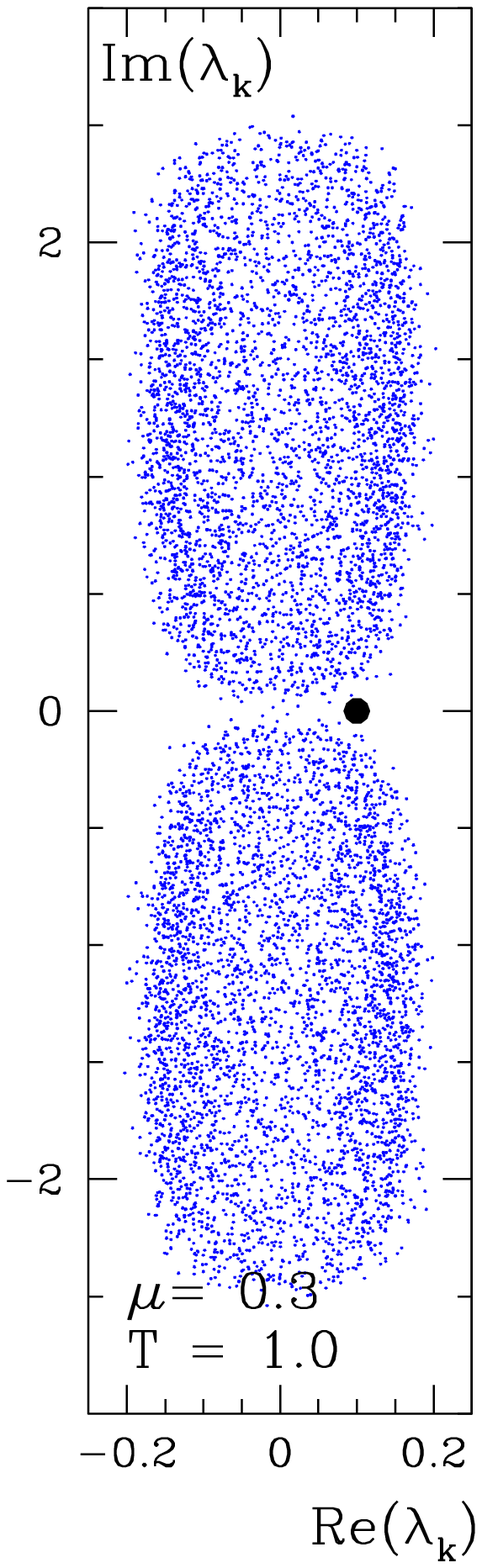,clip=,width=10cm}\hspace*{-6.5cm}
    \epsfig{file=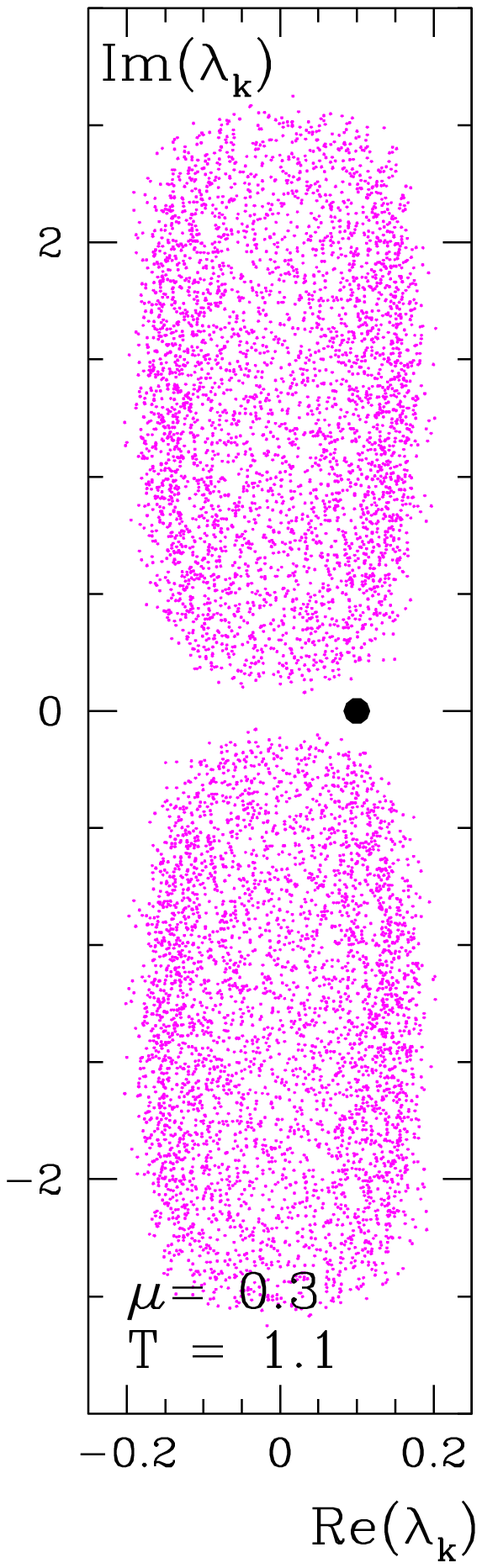,clip=,width=10cm}
  \caption{
Scatter plot of the eigenvalues of the random matrix Dirac operator for $\mu$ =0.3 and 
$T=0, \, 0.7, \, 0.9, \, 1.0, \, 1.1$ .
The quark mass at $m = 0.1$ is indicated by the black dot.}
  \label{fig4sc}
\end{center} 
\end{figure}

The results of the previous two sections show that 
the sign problem is
severe for QCD in one dimension with gauge group $U(N_c)$ and $\mu > \mu_c$.
Although this theory
does not have a critical temperature, it is clear from Fig. 1 that 
the average phase factor
for $\mu > \mu_c$ increases significantly for higher temperatures
(i.e. for lower values of $n$). However, this is a geometric effect
due to the Boltzmann factor.
Recent 
lattice simulations 
\cite{fodor1,fodor2,maria1d,maria,owe1,owe2,Allton,Allton2,Allton3,gupta,gupta2,schmidt,Ejiri,azco} 
suggest  
that the sign problem
becomes much milder for $T$ around $T_c$. Because of the absence of a critical
temperature, this temperature dependence of the average phase factor 
 cannot be investigated in one dimensional QCD. Instead
we study this problem in a schematic random matrix model at nonzero temperature
and chemical potential \cite{JV,misha,HJV}. This model has a severe sign problem
as well as a critical temperature and its eigenvalues are scattered in a finite
domain of the complex plane, just like in QCD. The quenched
average phase factor in this model is given by the matrix integral
\be
\langle e^{2i\theta} \rangle(m,\mu,T)= \int dC e^{-N{\rm Tr} C C^\dagger}
\frac{ \det (D+m)}{\det(D^\dagger +m)},
\label{thrmt}
\ee
with Dirac operator given by
\be
D= \mat 0 & iC + \mu +it \\ iC^\dagger + \mu +it &0\emat,
\label{rmtd}
\ee
and $t$ is the traceless diagonal matrix
$t= {\rm diag}( -T,\cdots,-T,T,\cdots, T)$. The integral is over the 
real and imaginary parts of the matrix elements of the complex $N\times N$ matrices $C$.
The phase quenched average phase factor is defined by
\be
\langle e^{2i\theta}\rangle_{\rm pq} =
\frac{ \int dC {\det}^2 (D+m) e^{-N{\rm Tr} C C^\dagger}}
{ \int dC {\det} (D+m) \det(D^\dagger +m)e^{-N{\rm Tr} C C^\dagger}}
\label{thrmtpq}
\ee
\begin{figure}[t!]
\begin{center}
  \unitlength1.0cm  
    \epsfig{file=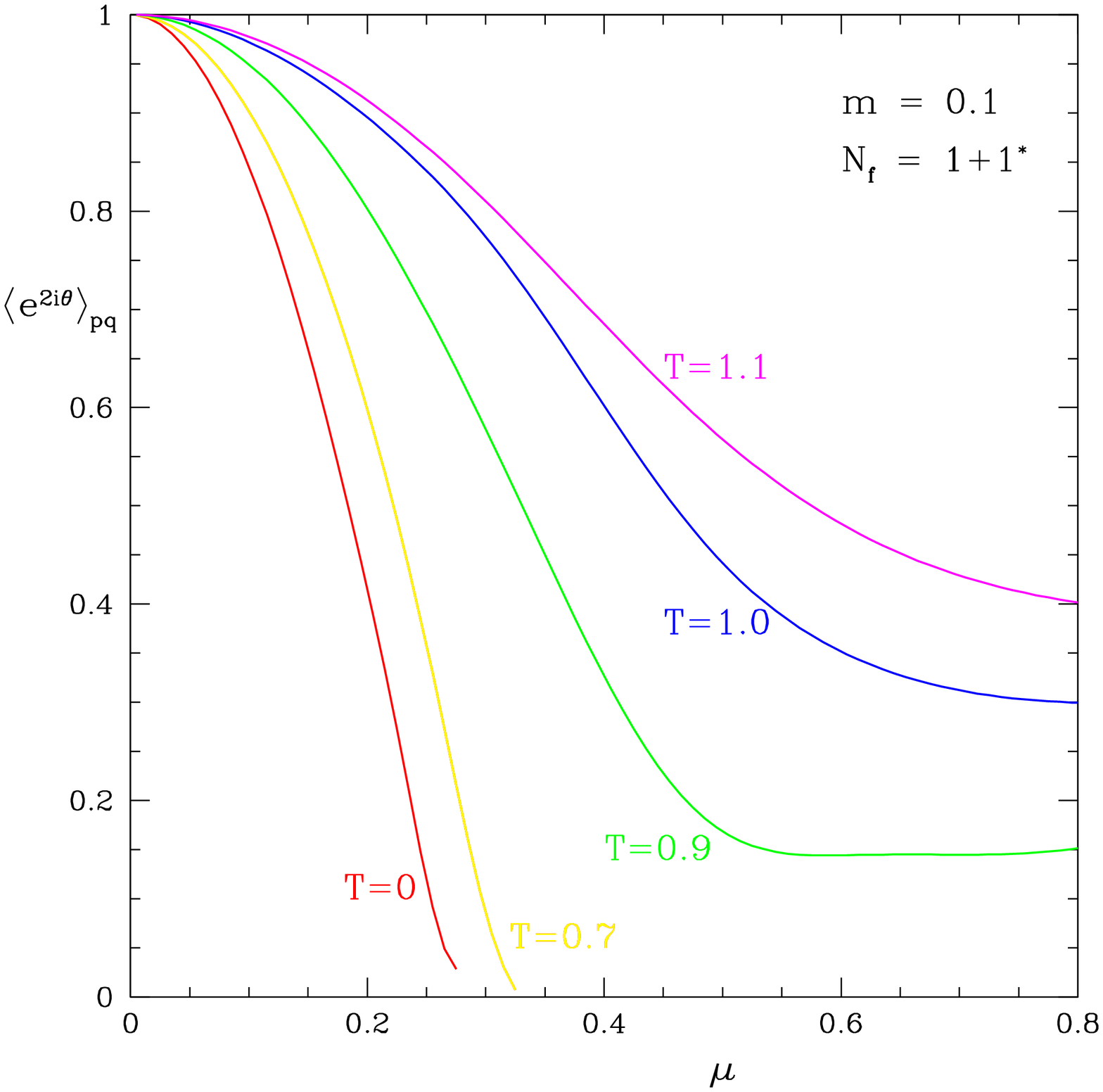,clip=,width=8cm}
    \epsfig{file=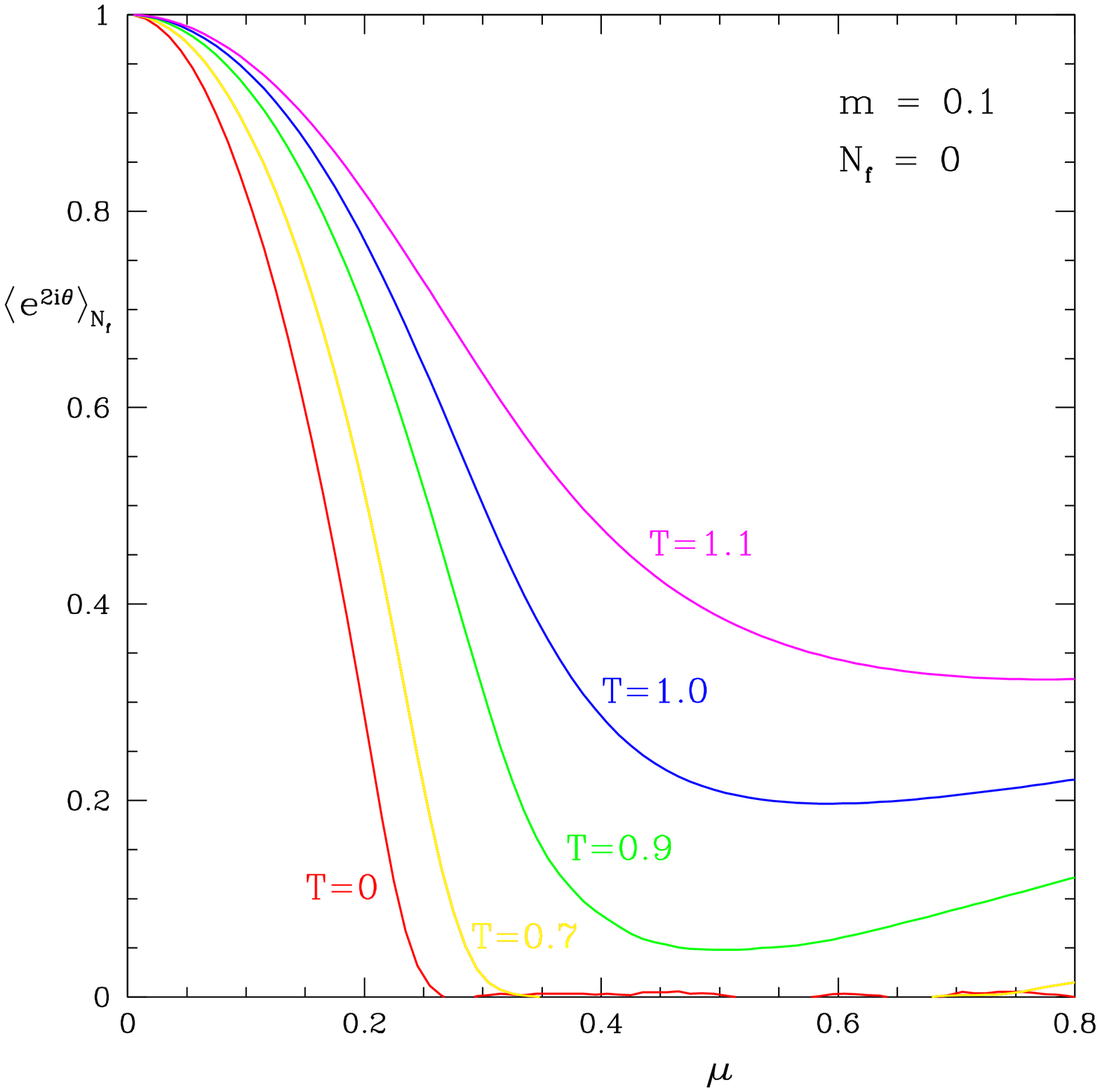,clip=,width=8cm}
  \caption{
Phase quenched (left) and quenched (right) 
average phase factor versus chemical potential. Results have been obtained
from the chiral random matrix model (\ref{rmtd}).}
  \label{fig4a}
\end{center} 
\end{figure}

If at all possible, it certainly
takes a significant effort to evaluate these integrals analytically. Therefore
we have studied the quenched (\ref{thrmt}) and phase quenched
(\ref{thrmtpq}) average phase factor  numerically. In Fig. 5, we
give scatter plots of the eigenvalues of the random matrix Dirac operator. For 
each of the temperatures
$T = 0, \, 0.7, \, 0.9, \, 1, \, 1.1$, the figures show results of 40 matrices
with $N=100$ and $\mu=0.3$. The black dot represents the quark 
mass for which the 
average phase factor is calculated.
In Fig. 6  we show the $\mu$-dependence of the phase quenched (left) and
quenched (right) average phase factor
obtained by averaging over 10,000 to 80,000 matrices $C$
with $N= 100$. The results for the phase quenched average phase factor 
for $T=0$ and $T=0.7$ and $\mu > \mu_c$ are not displayed because they
do not converge
because of the severity of the sign problem.
The average phase factor depends only weakly on 
the size of the matrices.
The conclusion of Figs. 5 and 6 is that the sign problem is not serious when
the quark mass is outside the domain of the eigenvalues. This 
confirms the conjecture that was made in \cite{kimcom,SVphase1,SVphase2}. 
It also agrees
with the observation in \cite{misha} that the free energy of the full theory
and the phase quenched theory are the same for $T=0$  
if the quark mass is outside the domain
of the eigenvalues (this is the case for $\mu<\mu_c$). The $\mu$-dependence
is due to the curvatures at the saddle point.

\section{Comparison with the QCD$_3$ Partition Function}
\label{qcd3} 

According to the general arguments given in section II~B, the microscopic
limit of the one dimensional QCD partition function at nonzero chemical
potential is equal to 
the microscopic
limit of the QCD$_3$ partition function at zero chemical potential, but
with shifted quark masses. In this section we show this explicitly for
some of the results derived before. We first discuss the microscopic limit
of the QCD$_3$ partition function which  can be  derived from 
either a chiral Lagrangian or 
a random
matrix model with the same global symmetries \cite{jvzahed3}. For a discussion
of QCD$_3$ in terms of chiral Lagrangians we refer to \cite{dunne}.

\subsection{Random Matrix Model}

From the continuum Dirac operator (\ref{dirac1d}) it follows  that
the random matrix model that describes the fluctuations of the low-lying Dirac eigenvalues is given by
\be
Z = \int dH P(H) \det ( iH +\mu +m) \det( iH -m +\mu),
\label{zr3}
\ee
where the probability distribution can conveniently be taken to be the
Gaussian distribution
\be
P(H) = e^{-2N \Sigma^2 {\rm Tr} H^2}.
\ee

Such random matrix partition functions have been studied 
elaborately in the literature 
to analyze the microscopic limit of QCD in three dimensions
\cite{jvzahed3,DN,AD3,ADDV,rep3}.
The microscopic limit of the partition 
function (\ref{zr3}) with quark masses given by 
$M={\rm diag}(-m_1,\cdots,-m_{N_f},m_1,\cdots, m_{N_f})$ is equal to \cite{DN,ADDV}
\be
Z_{\rm QCD_3} = \frac 1{\Delta(M)} \det \mat A(m)& A(-m) \\ A(-m) & A(m) \emat,
\label{zqcd3}
\ee
where the  matrix elements of the $N_f \times N_f$ matrix $A(m)$ are given by
\be
A(m)_{kl} = m_k^l e^{-m_k},
\ee
and $\Delta$ is the Vandermonde determinant 
\be
\Delta(M) = \prod_{k>l}^{2N_f} (M_{kk}-M_{ll}).
\ee

Although general expressions for QCD$_3$ partition functions with an 
arbitrary number of bosonic and fermionic determinants are also 
known \cite{akfyo}, for
our purposes we only need the partition function
\be
Z_{N_f+2|2} = \left \langle  \prod_{k=1}^{N_f}\det (D +m_k) \det(D-m_k)
\frac{\det (D+z_1) \det(D-z_2)  }{\det (D+\bar z_1) \det(D- \bar z_2)} 
\right \rangle .
\label{super3}
\ee
This partition function was evaluated in \cite{szabo} by means of the supersymmetric method 
and is given by
\be
Z_{N_f+2|2}&=&\frac{(z_1-\bar z_1)(z_2-\bar z_2)}{\Delta^2_{N_f}(m)}
\frac{\exp{N_c(\bar z_1+\bar z_2)}}{N_c(\bar z_1 + \bar z_2)}
\prod_{k=1}^{N_f} \frac{(m_k - \bar z_1)(m_k - \bar z_2)}{(m_k-z_1)(m_k-z_2)}
\det \left [ \begin{array}{cc}
\frac{\sinh N_c(m_k+m_l)}{N_c(m_k+m_l)}&\frac{\sinh N_c(m_k+z_1)}{N_c(m_k+z_1)}\\
\frac{\sinh N_c(z_2+m_l)}{N_c(z_2+m_l)}&\frac{\sinh N_c(z_1+z_2)}{N_c(z_1+z_2)}
\end{array} \right ]\nn\\
&&+ \frac {e^{-N_c(z_1-\bar z_1)+N_c(z_2-\bar z_2)}}{\Delta^2_{N_f}(m)}
\det\left [ \frac{\sinh N_c(m_k+m_l)}{N_c(m_k+m_l)}\right ].
\label{z-szabo}
\ee
The last term is a so-called Efetov-Wegner term, and the Vandermonde
determinant is over positive masses only
\be
\Delta(m) = \prod_{k>l}^{N_f} (m_k-m_l).
\ee

 \subsection{Microscopic Limit}

We will now show that the microscopic limit of the partition function 
(\ref{conreyform}) is equal to
the microscopic limit of the QCD$_3$ partition function. To this end we multiply
row $k$ of the first $N_f$ rows 
of the determinant by $\exp(-N_c m_{-\, k}/2)$ and row $k$ of the 
second $N_f$ rows of the determinant in (\ref{conreyform})
by $\exp(-N_c m_{+\, k}/2)$.
In the microscopic limit we keep $N_cm_{\pm,k}$ fixed so that we
can expand the masses that do not occur in this combination. By
subtracting successive columns starting with the first one we obtain
\be
\label{q3}
 Z_{N_f}&=& \frac {1}{\prod_{1\le k<l\le 2N_f}(M_l-M_k)}
\\&&\times
\left | \begin{array}{ccccccc}
 e^{-N_c m_{-\,1}/2}  & m_{-\, 1} e^{-N_cm_{-\,1}/2}& \cdots 
& m_{-\, 1}^{N_f -1} e^{-N_cm_{-\,1}/2} &m_{-\, 1}^{N_f }e^{+N_cm_{-\,1}/2} &\cdots &
m_{-\, 1}^{2N_f-1}e^{N_c m_{-\,1}/2}\\
\vdots & \vdots & &\vdots & \vdots & & \vdots \\
e^{-N_c m_{-\,N_f}/2}  & m_{-\, N_f} e^{-N_cm_{-\,N_f}/2}& \cdots 
& m_{-\, N_f}^{N_f -1} e^{-N_cm_{-\,N_f}/2} &m_{-\, N_f}^{N_f }e^{+N_cm_{-\,N_f}/2} 
&\cdots &m_{-\, N_f}^{2N_f-1}e^{N_c m_{-\,N_f}/2}\\
e^{-N_c m_{+\,1}/2}  & m_{+\, 1} e^{-N_cm_{+\,1}/2}& \cdots 
& m_{+\, 1}^{N_f -1} e^{-N_cm_{+\,1}/2} &m_{+\, 1}^{N_f }e^{+N_cm_{+\,1}/2} &\cdots &
m_{+\, 1}^{2N_f-1}e^{N_c m_{+\,1}/2}\\
\vdots & \vdots & &\vdots & \vdots & & \vdots \\
e^{-N_c m_{+\,N_f}/2}  & m_{+\, N_f} e^{-N_cm_{+\,N_f}/2}& \cdots 
& m_{+\, N_f}^{N_f -1} e^{-N_cm_{+\,N_f}/2} &m_{+\, N_f}^{N_f }e^{+N_cm_{+\,N_f}/2} 
&\cdots &m_{+\, N_f}^{2N_f-1}e^{N_c m_{+\,N_f}/2}
\end{array} \right |,\nn
\ee
where the masses $M_K$ are defined below Eq. (\ref{conreyform}).
By multiplying the columns $l=N_f+1, \cdots, 2N_f$ by $(-1)^{l-1}$ and introducing
microscopic masses,
we obtain exactly
the expression for the QCD$_{3}$ partition function given in (\ref{zqcd3}).

For masses $-(\mu_c +\mu), \,-(\mu_c-\mu), \, \mu_c-\mu, \,  
\mu_c +\mu$ corresponding to the microscopic limit of the two-flavor phase quenched
partition function given in (\ref{zpqmicro}) the determinant in (\ref{q3}) is given by
\be
4n\mu^2(e^{2nN_c\mu_c}+e^{-2nN_c\mu_c})-4n\mu_c^2(e^{2nN_c\mu}+e^{-2nN_c\mu})
+8n(\mu_c^2-\mu^2).
\ee
and the prefactor is equal to
\be
\prod_{1\le k<l\le 2N_f}(M_l-M_k)= 64n^6(\mu_c^2-\mu^2)\mu_c^2\mu^2.
\ee
Their ratio coincides with the microscopic phase quenched partition function given in
(\ref{zpqmicro}).

As second example, we consider the quenched average phase factor for $U(N_c)$ given in (\ref{phfull}). For $N_f=0$ the partition function (\ref{z-szabo})
simplifies to
\be
Z_{2|2} = (z_1-\bar z_1)(z_2 - \bar z_2) \frac {e^{N_c(\bar z_1 +\bar z_2)}}
{N_c(\bar z_1 + \bar z_2)}\frac{\sinh N_c(z_1+z_2)}{N_c(z_1+z_2)} 
+e^{-N_c(z_1-\bar z_1-z_2+\bar z_2)}.
\label{z2s2}
\ee
The masses in this partition function corresponding to (\ref{phfull}) are 
given by
\be
z_1 = n\mu_c +n \mu, &\qquad& z_2 = n\mu_c - n\mu,\nn\\
\bar z_2 = n\mu - n\mu_c, &\qquad& \bar z_1 = -n\mu - n\mu_c.
\label{qmass}
\ee
 Substituting them into (\ref{z2s2})  we obtain
\be
Z_{2|2} = 1 -\frac{\mu^2}{\mu_c^2}  + \frac{\mu^2}{\mu_c^2} e^{-4N_cn\mu}
\ee
which is exactly the microscopic limit of the quenched average phase factor (\ref{phquen}) given in Eq. (\ref{phmicro3}).

\begin{center}
\begin{figure}[t!]
\includegraphics[width=9.cm]{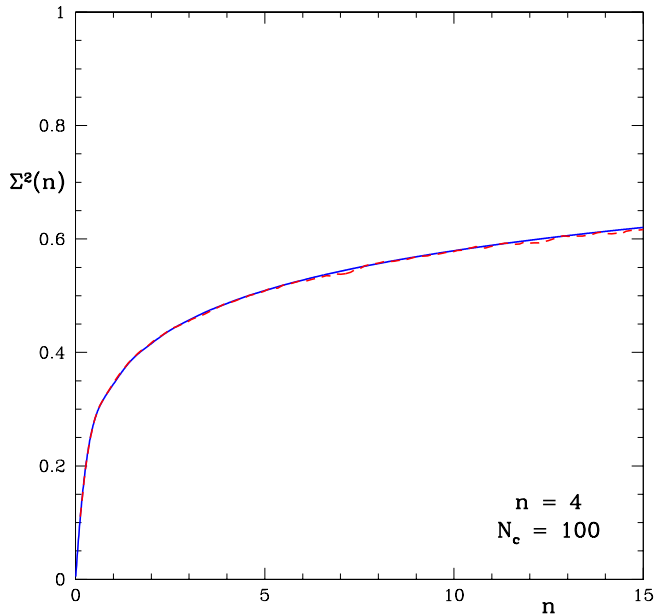}
\caption{The number variance for $n=4$ and $N_c = 100$.} 
\label{numvar}
\end{figure}
\end{center}

\section{Dirac Spectrum}

We consider the Dirac operator (\ref{dirac}) in a gauge where all gauge fields
except $U_{n1} \equiv U$ and $U_{1,n} = U^\dagger$  are equal to unity.
If the eigenvalues of  $U$ are equal to  $\exp(i\theta_k)$
the eigenvalues of the Dirac operator 
are given by
\be
\lambda_{k,l} = \frac 12(e^{\frac{2\pi i(k+1/2) + i\theta_l}n+\mu}
-e^{-\frac{2\pi i(k+1/2) - i\theta_l}n-\mu}),\quad k=1,\cdots, n, \quad
l = 1, \cdots, N_c.
\label{lambdak}
\ee
Contrary to QCD at $\mu \ne 0$ in more than one dimension,
where the eigenvalues are scattered in the complex plane 
\cite{all,misha,TV,SplitVerb2,tilo-ger,bloch},
the eigenvalues are located on an ellipse in the complex plane
with real and imaginary parts related by
\be
\left ( \frac{ {\rm Re}(\lambda_{k,l})}{e^\mu - e^{-\mu} }\right )^2
+\left ( \frac{ {\rm Im}(\lambda_{k,l})}{e^\mu + e^{-\mu}} \right )^2 =1.
\label{ellips}
\ee

\subsection{Universal Fluctuations}

In section \ref{sec:continuum} we have shown that the continuum limit of the staggered Dirac operator 
in one dimension is
in the same universality class as QCD in three dimensions. The
continuum limit of the staggered three dimensional Dirac operator, though,  is in the chiral symmetry universality
class of QCD in four dimensions. For example,  in \cite{dam3} this was found for the distribution of the 
small Dirac eigenvalues. 
The reason for the chiral structure is that the staggered Dirac operator
only couples  even and odd lattice sites. 
This is also the case in one dimension, but the off-diagonal
blocks, containing the gauge fields, 
are the same in the continuum limit (or occur in the combination
$U^n +U^{\dagger\, n}$), resulting in a two flavor theory with opposite masses (see (\ref{dirac3})).
The secular equation is given by
\be
\det[(\del_0+iA_0 + \mu+\lambda)(\del_0+iA_0+\mu - \lambda)] = 0.
\ee
This corresponds to the superposition of the spectrum   of $\del_0+A_0+\mu$ and $-\del_0-iA_0-\mu$. In the domain of the Dirac spectrum 
where $\del_0$ can be neglected, the eigenvalues on each of the lines $\pm \mu$ 
are given by the Hermitean random matrix ensemble $A_0$.
For $\mu =0$ we have the superposition of the Hermitean 
random matrix ensembles of $A_0$ and $-A_0$. 
From the eigenvalues (\ref{lambdak}) it is clear that
the staggered lattice Dirac eigenvalues show  a similar superposition. One ensemble is localized on the left half of the
ellipse and the other ensemble on its right half. Superpostions of more ensembles arise for correlations on
scales larger than $N_c$ level spacings.

Eigenvalues of $U(N_c)$ matrices are correlated according to the circular unitary ensemble. 
Therefore, the eigenvalues $\lambda_{k,l}$ are correlated according the 
Gaussian Unitary Ensemble (GUE)
on a scale where the variations in  the  average spectral density can be neglected, i.e. on a
scale much less than $N_c$ level spacings. By correcting for the variation of the 
average level spacing, the scale on which universal random matrix correlations are found can
be extended, but this scale should always remain well below $N_c$ level spacings.
Then we are necessarily  within one
of  the  $n$ successive copies of the eigenvalues along the ellipse. At a scale of  a
finite number of average level spacings, we expect convergence to universal random matrix
correlations in the large $N_c$ limit.

As measure of the correlations between  eigenvalues, we use the number variance $\Sigma^2(\bar n)$
defined as the variance on the number of eigenvalues in an interval that contains $\bar n$
 eigenvalues on average. We calculate the number variance for intervals starting from zero.
Because of the superposition of an enemble and its negative, this is equal to the number variance
of a single ensemble for an interval that is symmetric about zero which is given by the GUE result.
In Fig. \ref{numvar} we show the number variance obtained from an average over $10^5$
gauge field configurations for $N_c = 100$ and $\mu =0$. The number variance is calculated for an
interval starting at zero. The discrepancy between the analytical result and the GUE 
(solid curve)  is only barely visible. 

The increase of the domain of validity of chiral random matrix theory with increasing $N_c$
was also observed for Dirac operator of QCD in four dimensions
\cite{narayanan}. 
This can be explained as follows:
The number of eigenvalues that is described by random matrix
theory is equal to $\approx F_\pi^2 \sqrt V$ (with $V$ the space-time volume), but $F_\pi^2$
scales with  $N_c$ in the large $N_c$ limit.

\subsection{Chiral Symmetry Breaking at $\mu \ne 0$}

We first discuss the case of $\mu = 0$. Then
the average spacing of the eigenvalues scales as $1/(nN_c)$. This means
that  the chiral condensate develops a discontinuity at $m = 0$
when $nN_c \to \infty$. Two cases of interest where a nonzero chiral condensate can be obtained are 
the limit  $n \to \infty$ for fixed $N_c$,  i.e. at  zero temperature, or at
fixed $n$ in the limit $N_c \to \infty$. In both cases 
 the chiral condensate is discontinuous for $nN_c\to \infty$
when the quark mass crosses the imaginary axis where the eigenvalues
are located.
However, Goldstone bosons
and universal behavior is only found  in the large $N_c$ limit. At fixed $N_c$, chiral symmetry
breaking is not associated with universal behavior. 

For $\mu\ne 0$ the Dirac eigenvalues are located on the ellipse (\ref{ellips}).
 When the quark mass is inside the ellipse of eigenvalues the chiral condensate
is zero for $nN_c \to \infty$ for the $SU(N_c)$ partition function and 
the phase quenched $U(N_c)$ partition function \cite{bilic}. 
For gauge group $U(N_c)$ with $N_f\ge 1$ flavors, though,  the partition
function is $\mu$-independent so that the chiral condensate is also nonzero if the mass is
inside the eigenvalue ellipse. The Banks-Casher formula fails in this case which is known
as the ``Silver Blaze Problem'' \cite{cohen}. The resolution is the same as for QCD
in four dimensions \cite{OSV}:
the re-weighted eigenvalue distribution defined as
\be
\rho^{\rm full}(z)= \frac 1{Z_{N_f}}\int_{U(N_c)} dU {\det}^{N_f}D \sum_k \delta^2(z- \lambda_k)
\ee
shows oscillations with an amplitude that diverges exponentially with $n$ and a period that is 
proportional to $1/n$.
Below we will illustrate this for $U(1)$.

For $N_c =1$ one easily derives that
the spectral density is given by
\be
\rho^{\rm full}(z) = 4\frac{e^{n\mu_c} +e^{-n\mu_c} - e^{n(i\alpha+\mu)}- e^{-n(i\alpha+\mu)}}
{(e^{n\mu_c} +e^{-n\mu_c})(e^{2r} +e^{-2r} +e^{2i\alpha} + e^{-2i\alpha})} \delta(r-\mu),
\ee
where $z$ is parameterized as
\be
z = \frac 12 (e^{r+i\alpha} - e^{-r-i\alpha}),\quad r>0, \; \alpha \in [0,2\pi],
\label{zalr}
\ee 
and we have used that $Z_{N_f=1} = 2\cosh(n\mu_c)$ (see eq. (\ref{eq8})).
For $\mu> \mu_c$ this spectral density has oscillations with an amplitude that diverges 
exponentially with $n$ and a period of order $1/n$ which are the 
essential properties
of the spectral density of the Dirac operator of QCD in four dimensions
\cite{O,AOSV,OSV}.
It can be decomposed as
\be
\rho^{\rm full}(z)=\rho^{\rm q}(z) +\rho^{\rm osc}(z).
\label{decom}
\ee
The average quenched spectral density is defined by
\be
\rho^{\rm q}(z) = \frac 1{2\pi} \int d\theta \frac 1n \sum_{k=1}^n \delta^2(z - \lambda_k) 
\ee
with the eigenvalues $\lambda_k$ given by  Eq. (\ref{lambdak}).
After changing variables according to (\ref{zalr}) and integrating over $\theta$ we 
find
\be
\rho^{\rm q}(z) = \frac 1{2\pi} \frac 4
{(e^{2r} +e^{-2r} +e^{2i\alpha} + e^{-2i\alpha})} \delta(r-\mu).
\ee
The oscillatory part of the spectral density, $\rho^{\rm osc}(z)$  is equal  to the difference 
$\rho^{\rm full}(z) - \rho^{\rm q}(z)$.

The chiral condensate is given  by
\be
\Sigma^{\rm full}(m) = \int d^2z \frac{\rho^{\rm full} (z)}{z+m},
\ee
and can also be decomposed as
\be
\Sigma^{\rm full}(m) = \Sigma^{\rm q}(m) +\Sigma^{\rm osc}(m).
\ee  
The Jacobian for the transformation of $d^2z$ to $dr d\alpha$ is given by
\be
d^2z = \frac 14(e^{2 r} +e^{-2r } +e^{2i\alpha} + e^{-2i\alpha})dr d\alpha,
\ee
so that the chiral condensate after integration over $r$ can be simplified to
\be
\Sigma^{\rm full}(m)=\frac 1\pi \int_0^{2\pi} d \alpha
\frac{e^{n\mu_c} +e^{-n\mu_c} - e^{n(i\alpha+\mu)}- e^{-n(i\alpha+\mu)}}
{(e^{n\mu_c} +e^{-n\mu_c})( e^{\mu+i\alpha} - e^{-\mu-i\alpha} -2m)} = 
\frac {\tanh(n\mu_c)}{\cosh \mu_c} .
\ee
This result remains finite for $n \to \infty$.
We remind the reader that the mass is parameterized as $m = \sinh \mu_c$.
This expression represents 
the resolvent at the quark mass which was evaluated in \cite{HJV} and is in agreement with 
earlier work \cite{gibbs1,bilic}. Decomposing the spectral density and the chiral condensate
according to (\ref{decom}) we obtain
\be
\Sigma^{\rm osc}(m) &=&  \theta(\sinh(\mu) -m)\, \Sigma^{\rm full}(m),\nn\\
\Sigma^{\rm q}(m) &=& \theta(m-\sinh(\mu))\,\Sigma^{\rm full}(m) ,
\ee
so that when the quark mass is inside the ellipse of eigenvalues, the entire chiral condensate
is due to the oscillatory part of the spectral density. 
The discontinuity in the chiral condensate is reminiscent 
to a Stokes phenomenon.
The alternative to the Banks-Casher relation
proposed in \cite{OSV} is also at work for QCD in one dimension. This solves 
the ``Silver Blaze Problem'' \cite{cohen}.

\section{Conclusions}

We have studied QCD in one dimension at nonzero chemical potential. Both
the full theory, its quenched and phase quenched versions and gauge
groups $U(N_c)$ and $SU(N_c)$ have been considered. In one dimension,
the QCD or QCD-like partition functions can be reduced to a single matrix
integral, which because of recent advances by Conrey, Farmer and Zirnbauer,
could be evaluated analytically. In this paper we have analyzed the small
mass behavior of the partition function, the nature of the sign problem,
and the relation between the Dirac spectrum and chiral symmetry breaking.

To put our results in perspective, we emphasize that 
QCD in one dimension, is quite different
from QCD in more dimensions. In particular, we wish to mention the 
following three points.
First, instead of being scattered in the
complex plane, the eigenvalues of the staggered Dirac operator
are located on an ellipse  in the complex plane. Second, phase 
transitions can only take place  for zero temperature or for $N_c \to \infty$.
At the critical chemical potential, a transition from the vacuum state to
a state of free quarks takes place. Third, color singlets are made out
of noninteracting quarks so that the critical chemical potential is
given by the quark mass. In particular, the critical chemical potential
for the meson state and the baryon state is the same.

Because the Dirac eigenvalues are located on a curve, the large $N_c$ limit of 
staggered lattice QCD in one
dimension is in the same chiral symmetry class as QCD in three dimensions.
We have shown this both by an explicit evaluation of the partition function,
and by analyzing the fluctuations  of the Dirac eigenvalues for large $N_c$.
We have used this equivalence to explain the behavior of the partition 
functions and the average phase factor in the microscopic limit.

Contrary to QCD in four dimensions, the sign problem for QCD in one dimension
is not severe in the case of gauge group $SU(N_c)$. One reason is that the
critical chemical potential for full QCD and phase quenched QCD is the same,
so that the parameter domain $m_\pi/2 < \mu < m_N/3$, where the sign problem 
becomes severe in four dimensions, is absent in one dimension. A second
reason is that for $\mu>\mu_c$ both the full theory and the phase quenched
theory become a theory of free quarks with the same free energy in the 
thermodynamic limit. 

For gauge group $U(N_c)$, on the other hand, the
sign problem is severe when $\mu>m_\pi/2$ both in one dimension and in four dimensions. The reason
is that the  $U(N_c)$ theory does not have charged excitations, whereas the
phase quenched theory has charged mesons resulting in a phase transition
at $\mu= m_\pi/2$.
We have  evaluated the average phase factor both by averaging with respect 
to the phase quenched partition function and the full partition function, and similar
conclusions have been reached. 

The condititon 
$\mu>m_\pi/2$ for having a severe sign problem
can be rephrased as the quark mass being inside the ellipse of eigenvalues. In more
dimensions this condition is that 
the quark mass is inside the support of the Dirac spectrum.
It also applies to nonzero temperature as we have demonstrated explicitly 
in the framework of a chiral random matrix model.

Also for $U(N_c)$ QCD in one dimension, the chiral condensate
is discontinuous across the imaginary axis in spite of the fact that there
are no Dirac eigenvalues. This can only mean
 that the phase of the fermion determinant is responsible for the
discontinuity. This  is what happens in four dimensions where
the discontinuity is due to a contribution to the spectral 
density that oscillates with a period of the inverse volume
and amplitude that diverges exponentially with the volume. 
 Exactly the same
mechanism, where  the discontinuity in the chiral condensate 
arises due to a Stokes like phenomenon,
is at work in one dimension. This suggests that this is a universal
mechanism for  theories with a sign problem. 

We end by repeating that QCD in one dimension with $SU(N_c)$ as gauge
group has no serious sign problem. Our hope is that part of this 
conclusion translates to four dimensions 
ameliorating  the sign problem for $\mu> M_N/3$.

\vspace{4mm}

\noindent
{\sl Acknowledgments.} 
We wish to thank K. Splittorff for numerous comments and criticism and
P.H. Damgaard, H. Neuberger and R.Pisarski are thanked
for valuable discussions. K. Splittorff is also thanked for a careful
reading of the manuscript.
This work was
supported  by 
U.S. DOE Grant No. DE-FG-88ER40388 (JV),
the Angelo Della Riccia Foundation (LR),
the Villum Kann Rassmussen Foundation (JV) and
the Danish National Bank (JV).

\appendix
\section{Average $U(N_c)$ Phase Factor for $N_f = 1$}

In this appendix we give explicit results for the average phase factor for $N_f =1$
defined by
\be
\langle e^{2i\theta}\rangle_{N_f=1} = \frac{Z_{2|1^*}(\mu_c,\mu) }
{Z_{N_f=1}(\mu_c,\mu) }.
\ee
The partition function $Z_{N_f=1}(\mu_c,\mu)$  was already  given in Eq. (\ref{eq8}). The numerator
can be obtained from  the CFZ formula.
 After
taking the limit of degenerate critical chemical potentials at the
end of the calculation, we obtain for $\mu < \mu_c$
\be
Z_{2|1^*}(\mu_c,\mu) &=& e^{n(N_c+1)\mu_c}e^{-4n\mu}
 \frac{(1-e^{2n(\mu+\mu_c)})^2(1-e^{2n(\mu-\mu_c)})^2}
{(e^{n\mu_c}-e^{-n\mu_c})^5}+e^{-3n(N_c+1)\mu_c}
\frac{(e^{n\mu}- e^{-n\mu})^4}{(e^{n\mu_c}-e^{-n\mu_c})^5}\nn\\
&&+e^{-n(N_c+1)\mu_c}[f_0+N_cf_1 +N_c^2 f_2].
\ee
 with
\be
f_0 &=&\frac{2e^{2n(\mu+\mu_c)}+2e^{2n(\mu-\mu_c)}-e^{4n\mu}-e^{-4n\mu} +4e^{2n\mu} 
+4e^{-2n\mu}-e^{4n(\mu-\mu_c)} -e^{4n(\mu+\mu_c)} +2e^{6n\mu_c-2n\mu} +
2e^{6n\mu_c+2n\mu}}{(e^{n\mu_c} - e^{-n\mu_c})^5}\nn\\
 && +\frac{-10e^{2n\mu_c}+6e^{4n\mu_c}-2 -5 e^{6n\mu_c} -e^{-2n\mu_c}  }
{(e^{n\mu_c} - e^{-n\mu_c})^5},
\nn\\
f_1 &=& \frac{(1-e^{-2n\mu})^2(3e^{2n\mu_c}+e^{-2n\mu_c}-2e^{2n\mu}-2e^{-2n\mu})}
{(e^{n\mu_c} - e^{-n\mu_c})^3},
\nn\\
f_2&=& -\frac{(1-e^{-2n\mu})^2(1-e^{n(\mu+\mu_c}))(1-e^{n(\mu-\mu_c}))}
{(e^{n\mu_c} - e^{-n\mu_c})^3}.
\ee
For $\mu > \mu_c $ a similar calculation  results in
\be 
Z_{2|1^*}= N_c e^{ n N_c(\mu_c -2 \mu)}
\frac{(1-e^{-2n(\mu+\mu_c)})(1-e^{-2n\mu})}
{(1-e^{-2n\mu_c})^2}
( 1+ \frac{g_1}{N_c})
+ N_ce^{-nN_c(2\mu+\mu_c)}\frac{(1-e^{2n(\mu_c-\mu)})(1-e^{-2n\mu})}
{(1-e^{2n\mu_c})^2}(1+\frac{g_2}{N_c})\nn \\
\ee
with
\be
g_1 &=& 1 +\frac 1{e^{2n\mu}-1}+\frac 1{e^{2n(\mu+\mu_c)}-1}
-\frac 2{e^{2n\mu_c}-1},\nn \\
g_2 &=& 1+ \frac{1}{e^{2n(\mu-\mu_c)}-1}+
\frac 1{e^{2n\mu}-1} -\frac 2{e^{-2n\mu_c}-1}.
\ee
We have checked that the results in this appendix  agree with the large $N_c$
limit of the average phase factor given in section IV B.



\begin{thebibliography}{99}


\bibitem{mariarev}
  M.~P.~Lombardo,
  arXiv:hep-lat/0612017.





\bibitem{schmidtrev}
  C.~Schmidt,
  PoS {\bf LAT2006}, 021 (2006)
  [arXiv:hep-lat/0610116].

\bibitem{kimrev}
  K.~Splittorff,
  arXiv:hep-lat/0610072.


\bibitem{SVphase1}
  K.~Splittorff and J.~J.~M.~Verbaarschot,
  [arXiv:hep-lat/0702011].

\bibitem{SVphase2}
  K.~Splittorff and J.~J.~M.~Verbaarschot,
  Phys.\ Rev.\ Lett.\  {\bf 98}, 031601 (2007)
  [arXiv:hep-lat/0609076].


\bibitem{gibbs1}
  P.~E.~Gibbs, Preprint PRINT-86-0389-GLASGOW, 1986;
  P.~E.~Gibbs,
  Phys.\ Lett.\ B {\bf 182} (1986) 369.

\bibitem{Toussaint}
  D.~Toussaint,
  Nucl.\ Phys.\ Proc.\ Suppl.\  {\bf 17}, 248 (1990).

\bibitem{deFL}
  P.~de Forcrand and V.~Laliena,
  Phys.\ Rev.\ D {\bf 61}, 034502 (2000)
  [arXiv:hep-lat/9907004].

\bibitem{NakamuraPhase}
  Y.~Sasai, A.~Nakamura and T.~Takaishi,
  Nucl.\ Phys.\ Proc.\ Suppl.\  {\bf 129}, 539 (2004)
  [arXiv:hep-lat/0310046].

\bibitem{Ejiri}
  S.~Ejiri,
  Phys.\ Rev.\ D {\bf 69}, 094506 (2004)
  [arXiv:hep-lat/0401012];
  S.~Ejiri,
  Phys.\ Rev.\ D {\bf 73}, 054502 (2006)
   [arXiv:hep-lat/0506023].

  \bibitem{schmidt} C.~Schmidt, Z.~Fodor and S.~D.~Katz,
  arXiv:hep-lat/0512032.

\bibitem{Allton}
  C.~R.~Allton {\it et al.},
  Phys.\ Rev.\ D {\bf 66}, 074507 (2002)
  [arXiv:hep-lat/0204010];
  C.~R.~Allton, S.~Ejiri, S.~J.~Hands, O.~Kaczmarek, F.~Karsch, E.~Laermann and C.~Schmidt,
  Phys.\ Rev.\ D {\bf 68}, 014507 (2003)
  [arXiv:hep-lat/0305007];
  C.~R.~Allton {\it et al.},
  Phys.\ Rev.\ D {\bf 71}, 054508 (2005)
  [arXiv:hep-lat/0501030].




\bibitem{SV}           
  E.~V.~Shuryak and J.~J.~M.~Verbaarschot,
  Nucl.\ Phys.\ A {\bf 560}, 306 (1993)
  [arXiv:hep-th/9212088].


\bibitem{V}                 
  J.~J.~M.~Verbaarschot, 
  Phys.\ Rev.\ Lett.\  {\bf 72}, 2531 (1994)
  [arXiv:hep-th/9401059].



\bibitem{CFZ} J.B. Conrey, D.W. Farmer and  M.R. Zirnbauer,
[arXiv:math-ph/0511024], 2005.

\bibitem{ilgenfritz}
  E.~M.~Ilgenfritz and J.~Kripfganz,
  Z.\ Phys.\  C {\bf 29}, 79 (1985).


\bibitem{damgaard}
  P.~H.~Damgaard, N.~Kawamoto and K.~Shigemoto,
  Phys.\ Rev.\ Lett.\  {\bf 53} (1984) 2211;
  Nucl.\ Phys.\  B {\bf 264}, 1 (1986).


\bibitem{gocksch-ogilvie}
  A.~Gocksch and M.~Ogilvie,
  Phys.\ Rev.\ D {\bf 31} (1985) 877.

\bibitem{hochberg}
  P.~H.~Damgaard, D.~Hochberg and N.~Kawamoto,
  Phys.\ Lett.\ B {\bf 158} (1985) 239.
\bibitem{bilic-new}
  N.~Bilic, K.~Demeterfi and B.~Petersson,
  Nucl.\ Phys.\ B {\bf 377}, 651 (1992).



\bibitem{kim}
  P.~de Forcrand and S.~Kim,
  arXiv:hep-lat/0608012.




\bibitem{gibbs-plb}
  P.~E.~Gibbs,
  Phys.\ Lett.\  B {\bf 172}, 53 (1986).




\bibitem{gocksch}
  A.~Gocksch,
  Phys.\ Rev.\  D {\bf 37}, 1014 (1988).

\bibitem{bilic}
  N.~Bilic and K.~Demeterfi,
  Phys.\ Lett.\  B {\bf 212}, 83 (1988).


\bibitem{gupta1d}
  S.~Gupta,
  Phys.\ Lett.\  B {\bf 588}, 136 (2004)
  [arXiv:hep-lat/0307007].

\bibitem{maria1d}
  M.~P.~Lombardo,
  Nucl.\ Phys.\ Proc.\ Suppl.\  {\bf 83}, 375 (2000)
  [arXiv:hep-lat/9908006].



\bibitem{all}I.~Barbour, N.~E.~Behilil, E.~Dagotto, 
F.~Karsch, A.~Moreo, M.~Stone and H.~W.~Wyld,
  Nucl.\ Phys.\ B {\bf 275}, 296 (1986).






\bibitem{jvzahed3}
  J.~J.~M.~Verbaarschot and I.~Zahed,
  Phys.\ Rev.\ Lett.\  {\bf 73}, 2288 (1994)
  [arXiv:hep-th/9405005].

\bibitem{stone}
  A.~J.~McKane and M.~Stone,
  Annals Phys.\  {\bf 131}, 36 (1981).

\bibitem{misha}
  M.~A.~Stephanov,
  Phys.\ Rev.\ Lett.\  {\bf 76}, 4472 (1996)
  [arXiv:hep-lat/9604003].

\bibitem{OSV}
  J.~C.~Osborn, K.~Splittorff and J.~J.~M.~Verbaarschot,
  Phys.\ Rev.\ Lett.\  {\bf 94}, 202001 (2005)
  [arXiv:hep-th/0501210].

\bibitem{HK}
  P.~Hasenfratz and F.~Karsch,
  Phys.\ Lett.\ B {\bf 125}, 308 (1983).


\bibitem{color-flavor}M.R. Zirnbauer, 
[arXiv:chao-dyn/9609007].

\bibitem{slit}
  B.~Schlittgen and T.~Wettig,
  Nucl.\ Phys.\  B {\bf 632}, 155 (2002)
  [arXiv:hep-lat/0111039].


\bibitem{zsu}
  J.~Budczies, S.~Nonnenmacher, Y.~Shnir and M.~R.~Zirnbauer,
  Nucl.\ Phys.\  B {\bf 635}, 309 (2002)
  [arXiv:hep-lat/0112018].



\bibitem{snaith} J.B. Conrey, D.W. Farmer, J.P. Keating, M.O. Rubinstein
and N.C. Snaith, [arXiv:math-ph/0208007].

\bibitem{forrester} E.L. Basor and P.J. Forrester, Mathematische Nachrichten,
{\bf 170}, 5 (1194).
 



\bibitem{fodor1}
  Z.~Fodor and S.~D.~Katz,
  JHEP {\bf 0203}, 014 (2002)
  [arXiv:hep-lat/0106002].

\bibitem{fodor2}
  Z.~Fodor and S.~D.~Katz,
  JHEP {\bf 0404}, 050 (2004)
  [arXiv:hep-lat/0402006].


\bibitem{owe1}
  P.~de Forcrand and O.~Philipsen,
  Nucl.\ Phys.\ B {\bf 642}, 290 (2002)
  [arXiv:hep-lat/0205016].

\bibitem{owe2}
  P.~de Forcrand and O.~Philipsen,
  Nucl.\ Phys.\ B {\bf 673}, 170 (2003)
  [arXiv:hep-lat/0307020].

\bibitem{maria}
  M.~D'Elia and M.~P.~Lombardo,
  Phys.\ Rev.\ D {\bf 67}, 014505 (2003)
  [arXiv:hep-lat/0209146].

 \bibitem{Allton2}
  C.~R.~Allton, S.~Ejiri, S.~J.~Hands, O.~Kaczmarek, F.~Karsch, E.~Laermann and C.~Schmidt,
  Phys.\ Rev.\ D {\bf 68}, 014507 (2003)
  [arXiv:hep-lat/0305007].

\bibitem{Allton3}
  C.~R.~Allton {\it et al.},
  Phys.\ Rev.\ D {\bf 71}, 054508 (2005)
  [arXiv:hep-lat/0501030].


\bibitem{gupta}
  R.~V.~Gavai and S.~Gupta,
  Phys.\ Rev.\ D {\bf 68}, 034506 (2003)
  [arXiv:hep-lat/0303013].

\bibitem{gupta2}
  R.~V.~Gavai and S.~Gupta,
  Phys.\ Rev.\  D {\bf 71}, 114014 (2005)
  [arXiv:hep-lat/0412035].


\bibitem{azco}
 V.~Azcoiti, G.~di Carlo and A.~F.~Grillo,
  Phys.\ Rev.\ Lett.\  {\bf 65}, 2239 (1990);
 J.~Ambjorn, K.~N.~Anagnostopoulos, J.~Nishimura and J.~J.~M.~Verbaarschot,
  JHEP {\bf 0210}, 062 (2002)
  [arXiv:hep-lat/0208025].


\bibitem{JV}
  A.~D.~Jackson and J.~J.~M.~Verbaarschot,
  Phys.\ Rev.\  D {\bf 53}, 7223 (1996)
  [arXiv:hep-ph/9509324].

\bibitem{HJV}
  M.~A.~Halasz, A.~D.~Jackson and J.~J.~M.~Verbaarschot,
  Phys.\ Rev.\ D {\bf 56} (1997) 5140
  [arXiv:hep-lat/9703006].

\bibitem{kimcom}
  K.~Splittorff,
  arXiv:hep-lat/0505001.

\bibitem{dunne}
  G.~V.~Dunne and S.~M.~Nishigaki,
  Nucl.\ Phys.\ B {\bf 670}, 307 (2003) [arXiv:hep-ph/0306220].



\bibitem{AD3}
  G.~Akemann and P.~H.~Damgaard,
  Nucl.\ Phys.\  B {\bf 576}, 597 (2000)
  [arXiv:hep-th/9910190].

\bibitem{DN}
  P.~H.~Damgaard and S.~M.~Nishigaki,
  Phys.\ Rev.\ D {\bf 57}, 5299 (1998)
  [arXiv:hep-th/9711096].

\bibitem{ADDV}
  G.~Akemann, D.~Dalmazi, P.~H.~Damgaard and J.~J.~M.~Verbaarschot,
  Nucl.\ Phys.\ B {\bf 601}, 77 (2001)
  [arXiv:hep-th/0011072].



\bibitem{rep3}  
T.~Andersson, P.~H.~Damgaard and K.~Splittorff,
  Nucl.\ Phys.\  B {\bf 707}, 509 (2005)
  [arXiv:hep-th/0410163].

\bibitem{akfyo}
  G.~Akemann and Y.~V.~Fyodorov,
  Nucl.\ Phys.\  B {\bf 664}, 457 (2003)
  [arXiv:hep-th/0304095].




\bibitem{szabo}
  R.~J.~Szabo,
  Nucl.\ Phys.\ B {\bf 598}, 309 (2001)
  [arXiv:hep-th/0009237].

\bibitem{TV}
  D.~Toublan and J.~J.~M.~Verbaarschot,
  Int.\ J.\ Mod.\ Phys.\  B {\bf 15}, 1404 (2001)
  [arXiv:hep-th/0001110].


\bibitem{SplitVerb2}
  K.~Splittorff and J.~J.~M.~Verbaarschot,
  Nucl.\ Phys.\  B {\bf 683}, 467 (2004)
  [arXiv:hep-th/0310271].
\bibitem{tilo-ger}
  G.~Akemann and T.~Wettig,
  Phys.\ Rev.\ Lett.\  {\bf 92}, 102002 (2004)
  [Ibid.\  {\bf 96}, 029902 (2006)]
 [arXiv:hep-lat/0308003].

\bibitem{bloch}
  J.~Bloch and T.~Wettig,
  Phys.\ Rev.\ Lett.\  {\bf 97}, 012003 (2006)
  [arXiv:hep-lat/0604020].




\bibitem{dam3}
  P.~H.~Damgaard, U.~M.~Heller, A.~Krasnitz and T.~Madsen,
  Phys.\ Lett.\  B {\bf 440}, 129 (1998)
  [arXiv:hep-lat/9803012].


\bibitem{narayanan}
  R.~Narayanan and H.~Neuberger,
  Nucl.\ Phys.\  B {\bf 696}, 107 (2004)
  [arXiv:hep-lat/0405025].
 




\bibitem{cohen}
  T.~D.~Cohen,
  Phys.\ Rev.\ Lett.\  {\bf 91}, 222001 (2003)
  [arXiv:hep-ph/0307089].

\bibitem{O}
  J.~C.~Osborn,
  Phys.\ Rev.\ Lett.\  {\bf 93}, 222001 (2004)
  [arXiv:hep-th/0403131].

\bibitem{AOSV}
  G.~Akemann, J.~C.~Osborn, K.~Splittorff and J.~J.~M.~Verbaarschot,
  Nucl.\ Phys.\  B {\bf 712}, 287 (2005)
  [arXiv:hep-th/0411030].


\end{thebibliography}
\end{document}